%% file: ms.tex
\documentclass[12pt,a4paper,oneside]{book}

\usepackage{amsmath}
\usepackage{amssymb}
\usepackage{verbatim}
\usepackage{tcolorbox}
\usepackage{enumitem}
\usepackage[boxruled,linesnumbered]{algorithm2e}
\usepackage[parfill]{parskip}
\usepackage{pdfpages}
\usepackage{fancyhdr}
\usepackage{fancyvrb}
\usepackage{float}
\usepackage{multirow,tabularx}
\usepackage{graphicx}
\graphicspath{{./images/}}
\usepackage{pdflscape}
\usepackage{acronym}
\usepackage[hyphens]{url}

\usepackage{amsthm}
\usepackage{thmtools}
\theoremstyle{definition}
\newtheorem{definitionsec}{Definition}[section]

\usepackage[pdftex,
	pdfauthor={Massimo Nazaria},
	pdftitle={Simulation Based Formal Verification of Cyber-Physical Systems},
	pdfsubject={PhD Thesis},
	pdfkeywords={Computer Science, Model Based Design, Software Verification, Simulation Based Formal Verification},
	pdfproducer={Latex with hyperref},
	pdfcreator={pdflatex}]{hyperref}

\usepackage[titletoc]{appendix}

\begin{document}
	\frontmatter

	\hypersetup{pageanchor=false}
	\begin{titlepage}
	\includepdf[pages=-]{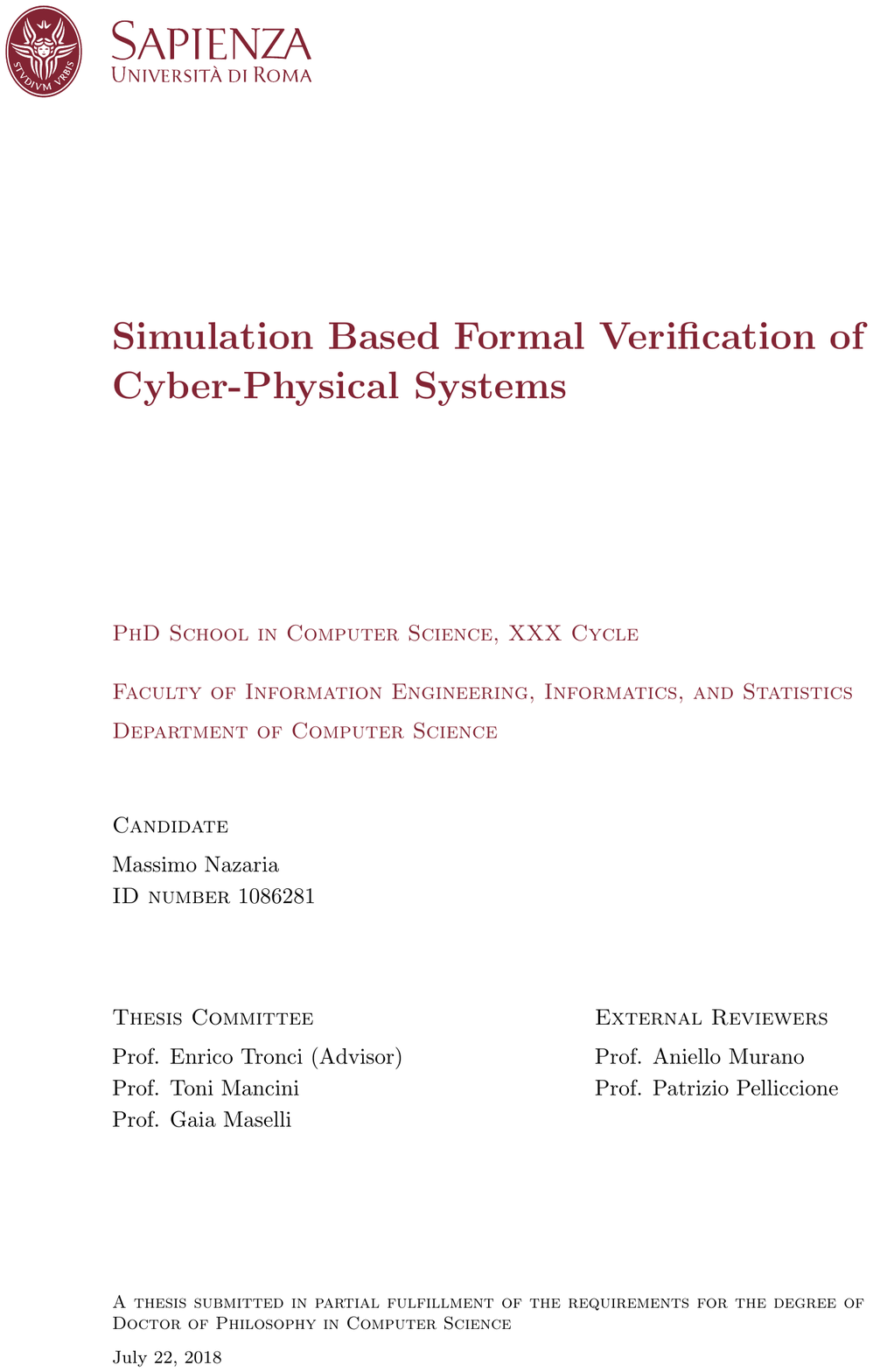}
	\end{titlepage}

	\input{author}

	\pagestyle{plain}
	\renewcommand{\thepage}{\roman{page}}

	\input{abstract}
	\input{dedication}
	\input{acknowledgements}
	\tableofcontents
	\clearpage
	\listoffigures
	\addcontentsline{toc}{chapter}{\listfigurename}
	\clearpage
	\listoftables
	\addcontentsline{toc}{chapter}{\listtablename}
	\clearpage
	\input{acronyms}

	\renewcommand{\thepage}{\arabic{page}}

	\mainmatter

	\input{chap01}
	\input{chap02}
	\input{chap03}
	\input{chap04}
	\input{chap05}

	\appendix
	\input{appendix}

	\backmatter

\input{bibliography}
\end{document}

%% file: author.tex
\thispagestyle{empty}

Copyright \textcopyright~2018 Massimo Nazaria\\
\vspace*{-8pt}\hrule
\vspace*{\fill}
\textsc{Author's Address}\\[5pt]
Massimo Nazaria\\[3pt]
Sapienza University of Rome\\[-1pt]
Department of Computer Science\\[-1pt]
Via Salaria 113, 00198 Rome, Italy\\[3pt]
{\sc Email:} \texttt{\href{mailto:nazaria@di.uniroma1.it}{nazaria@di.uniroma1.it}}

\clearpage

%% file: abstract.tex
\chapter*{Abstract}
\addcontentsline{toc}{chapter}{Abstract}

{\footnotesize
Cyber-Physical Systems (CPSs) have become an intrinsic part of the 21st century world.
Systems like Smart Grids, Transportation, and Healthcare help us run our lives and businesses smoothly, successfully and safely.
Since malfunctions in these CPSs can have serious, expensive, sometimes fatal consequences, System-Level Formal Verification (SLFV) tools are vital to minimise the likelihood of errors occurring during the development process and beyond.
Their applicability is supported by the increasingly widespread use of Model Based Design (MBD) tools.
MBD enables the simulation of CPS models in order to check for their correct behaviour from the very initial design phase.
The disadvantage is that SLFV for complex CPSs is an extremely
time-consuming process, which typically requires several months of simulation.
Current SLFV tools are aimed at accelerating the verification process with multiple simulators working simultaneously.
To this end, they compute all the scenarios in advance in such a way
as to split and simulate them in parallel.
Furthermore, they compute optimised simulation campaigns 
in order to simulate common prefixes of these scenarios only once, thus 
avoiding redundant simulation.
Nevertheless, there are still limitations that prevent a more widespread adoption of SLFV tools.
Firstly, current tools cannot optimise simulation campaigns from existing datasets with collected scenarios.
Secondly, there are currently no methods to predict the time required to complete the SLFV process.
This lack of ability to predict the length of the process makes scheduling verification activities highly problematic.
In this thesis, we present how we are able to overcome these limitations with the use of a simulation campaign optimiser and an execution time estimator.
The optimiser tool is aimed at speeding up the SLFV process by using a data-intensive algorithm to obtain optimised simulation campaigns from existing datasets, that may contain a large quantity of collected scenarios.
The estimator tool is able to accurately predict the execution time to simulate a given simulation campaign by using an effective machine-independent method.
}
\clearpage

%% file: dedication.tex
\thispagestyle{empty}

\vspace*{3cm}
\hfill{\it To Silvia}

\clearpage

%% file: acknowledgements.tex
\chapter*{Acknowledgements}
\addcontentsline{toc}{chapter}{Acknowledgements}

{\footnotesize
First of all, I would like to thank my advisor Professor Enrico Tronci for his precious
guidance and support over the past few years, and
for being an immense source of inspiration with his exceptional technical-scientific \mbox{knowledge} and
extraordinary human qualities.
Secondly, my thanks go to my colleagues at the Model Checking Laboratory\footnote{\url{http://mclab.di.uniroma1.it}} 
for their prompt help whenever I needed it, and for providing me with a friendly-yet-professional work environment.
Last but not least, I would like to thank my family and friends for their unconditional support, encouragement, and
love without which I would never have come this far.
}
\clearpage

%% file: acronyms.tex
\chapter*{List of Acronyms}
\addcontentsline{toc}{chapter}{List of Acronyms}

\begin{acronym}[TDMA]
	\acro{CPS}{Cyber-Physical System}
	\acro{CCP}{Coupon Collector's Problem}
	\acro{DFS}{Depth-First Search}
	\acro{DT}{Disturbance Trace}
	\acro{LBT}{Labels Branching Tree}
	\acro{LCP}{Longest Common Prefix}
	\acro{MBD}{Model Based Design}
	\acro{MCDS}{Model Checking Driven Simulation}
	\acro{ODE}{Ordinary Differential Equations}
	\acro{OP}{Omission Probability}
	\acro{PWLF}{Piecewise Linear Function}
	\acro{RMSE}{Root-Mean-Square Error}
	\acro{SBV}{Simulation Based Verification}
	\acro{SLFV}{System-Level Formal Verification}
	\acro{WCET}{Worst Case Execution Time}
\end{acronym}

\clearpage

%% file: chap01.tex
\chapter{Introduction}

\section{Framework}

Over the past few decades, we have experienced many technological breakthroughs which have revolutionised the way we live.
Cyber-Physical Systems (CPSs) such as smart grid, automotive, and medical systems contribute to ensuring our lives run smoothly,
safely and securely. They have become such an intrinsic part of our 21st century world,
that few people nowadays could imagine living without them.

Consequently, we have come to expect CPSs to autonomously cope with all the faulty events
originating from the environment they operate in.
For example, control software in a car must be able to detect any
hardware failure and take immediate mitigating action with
a high level of autonomy.

In order to achieve this, software components must ensure that all the relevant faulty events from the external environment are promptly detected, and appropriately dealt with in such a way as to either prevent system failures, or recover from temporary malfunctions in a reasonable amount of time.

Since malfunctions in these CPSs can have far-reaching, potentially dangerous, sometimes fatal consequences,
it is extremely important to use the most appropriate verification methods over the development life cycle.
For this reason, verification activities are performed in parallel with the development process, from the initial design up to the acceptance testing stages.

\clearpage
This is particularly the case at the earliest design stages, where both the cost and effort required
to fix design defects can be minimised.
Furthermore, the propagation of undiscovered design defects over subsequent
stages results in an exponential increase in the reworking needed to resolve them~\cite{Jones2000}.

In fact, the verification of automobile components must prove that the car works as expected in any operating conditions, including engine speed, road surface and ambient temperatures.
To this end, thousands of kilometers of road tests are typically needed to verify newly developed components on real vehicles.

Any bug found at this later verification stage has a huge impact on the project schedule and costs.
In order to mitigate this, proper simulation methods are used to carry out verification activities as early as possible.

\section[Model Based System-Level Formal Verification]{Model Based System-Level Formal\\Verification}

Verification methods and tools have long been established as the greatest contributing factor for any successful software development life cycle. Nowadays, there are a variety of verification tools available for all the specific development stages, including system-level design, unit-level coding, integration, and acceptance testing.

Model Based System-Level Formal Verification (SLFV) tools
are extremely valuable to fulfil the need to verify the correctness of
CPSs being developed during the earliest stages of
the development process.
One advantage of this model-based approach is that it brings verification forward in the development life cycle. Thus, developers are able to construct, test, and analyse their designs before any system component is implemented. In this way, if major problems are found, they can be resolved with less impact on the budget or schedule of the project.

The goal of SLFV is to use simulation to build confidence that
the system as a whole behaves as expected regardless of disturbances
originating from the operational environment.
Examples of disturbances include \textit{sensor }$x$\textit{ has a failure}, and \textit{parameter }$y$\textit{ has been changed}.
The fact that there are typically a huge number of disturbance scenarios that can originate from the environment makes SLFV of complex CPSs very time-consuming.

\clearpage
\subsubsection*{Simulation Based Verification}

\begin{figure}[H]
\begin{center}
	\includegraphics[width=1\linewidth]{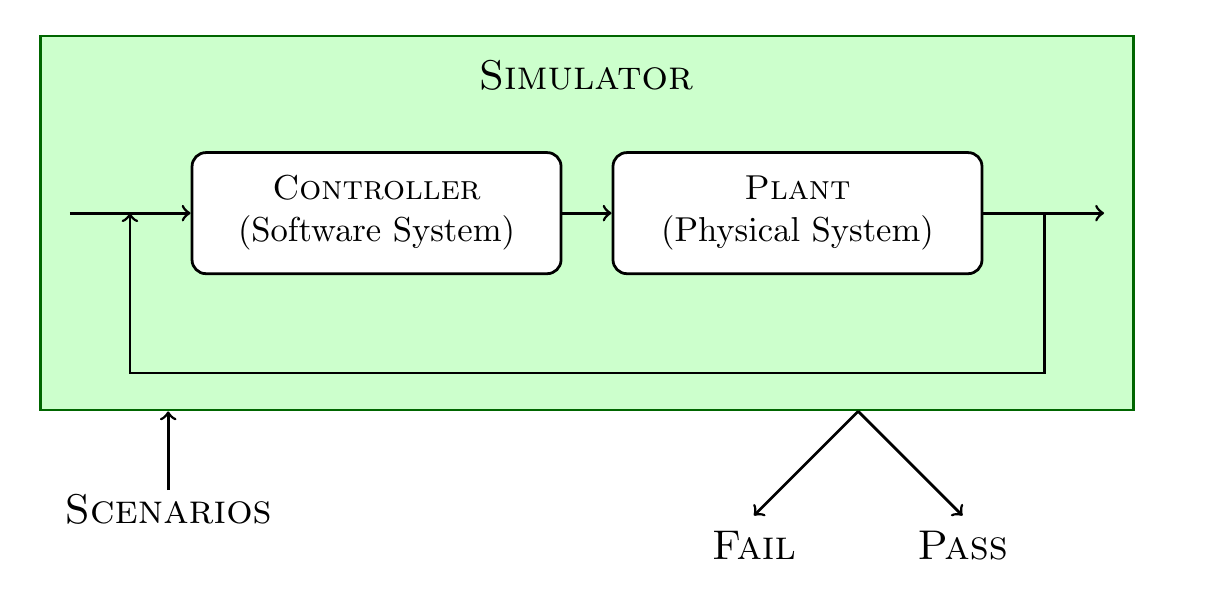}
	\caption{Simulation Based Verification\label{fig:intro:closedloop}}
\end{center}
\end{figure}

The most commonly used approach to carrying out SLFV is called
Simulation Based Verification (SBV), which consists of
simulating both hardware and software components of the CPS at the same time
in order to reproduce all the relevant operational scenarios.
As an example, Figure~\ref{fig:intro:closedloop} shows a
typical SBV setting.

The applicability of SLFV is supported by the increasing use of
Model Based Design (MBD) tools to develop CPSs.
This is particularly the case within the Automotive, Space, and High-Tech sectors,
where tools like Simulink have been established as \textit{de facto} standards.
Such a widespread adoption of MBD tools is mainly due to the fact that
they enable SLFV activities at an early stage in the development life cycle.

In this way, when designing a new feature for an existing CPS, modelling and simulation activities can be carried out in order to detect potential design defects that would negatively impact the total development effort and cost.
In fact, the re-working that would be required to resolve such defects increases exponentially,
if the defects are able to propagate undiscovered over subsequent development stages.

Thus, designers are able to analyse the behaviour of CPSs
before actual implementation begins.
Needless to say, detecting and fixing errors at the design stage implies 
significantly less cost and effort compared to later integration and 
acceptance testing stages.

Another advantage is clearly the potential for shorter time-to-market,
which is useful for producers to accelerate getting their product to market so as to maximise the volume of sales after development kickoff.

\subsubsection*{Simulation Based Verification Process}

\begin{landscape}
\begin{figure}[H]
\begin{center}
	\includegraphics[width=0.7\linewidth]{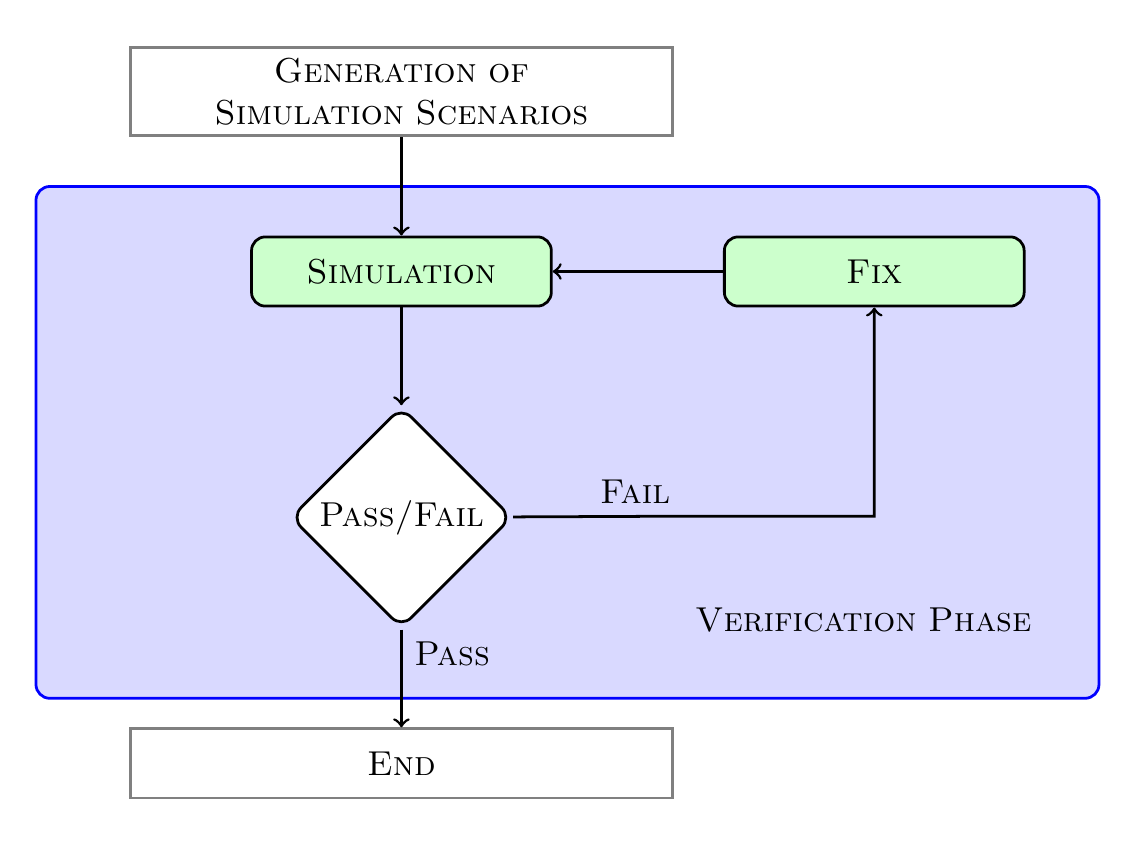}
	\caption{Simulation Based System-Level Verification Process\label{fig:intro:hils}}
\end{center}
\end{figure}
\end{landscape}

Figure~\ref{fig:intro:hils} illustrates a high-level diagram of a typical SBV process.
In a nutshell, it consists of the following two main phases.
First, the \textsc{Generation Phase} where simulation scenarios are generated.
Second, the \textsc{Verification Phase} where SBV is carried out.

In the case of failure, design errors are fixed and 
simulation is repeated until the final output is \textsc{Pass}.

\subsubsection*{Operational Environment and Safety Properties Specification}

SLFV uses formal languages to specify both the environment where the system
will operate, and the safety properties it is necessary to satisfy,
regardless of any faulty events originating in the
specified environment.
These formal specification languages include Simulink, and SysML\footnote{\url{http://sysml.org/}}.

As a result of these specifications, SLFV can be performed
using an assume-guarantee approach.
Namely, an SLFV output of \textsc{Pass} guarantees that
the system will behave as expected, assuming that 
all and only the relevant scenarios are formally specified.

\clearpage
\section{State of the Art}

Notwithstanding the significant advantages of using SLFV,
the verification of \mbox{complex} CPSs still remains
an extremely time-consuming process. As a matter of fact, SLFV can take
up to several months of simulation since it requires
a considerable amount of scenarios to be analysed~\cite{PDP2014}.

In order to mitigate this, techniques aimed at speeding up the SLFV process
have recently been studied. The two approaches described below use formal methods to specify both
the operational environment of the CPS and the safety properties to verify.

\subsection[SLFV via Parallel Model Checking Driven Simulation]{System-Level Formal Verification via Parallel\\Model Checking Driven Simulation}

\begin{landscape}
\begin{figure}[H]
\begin{center}
	\vspace*{30pt}
	\includegraphics[width=1.255\linewidth]{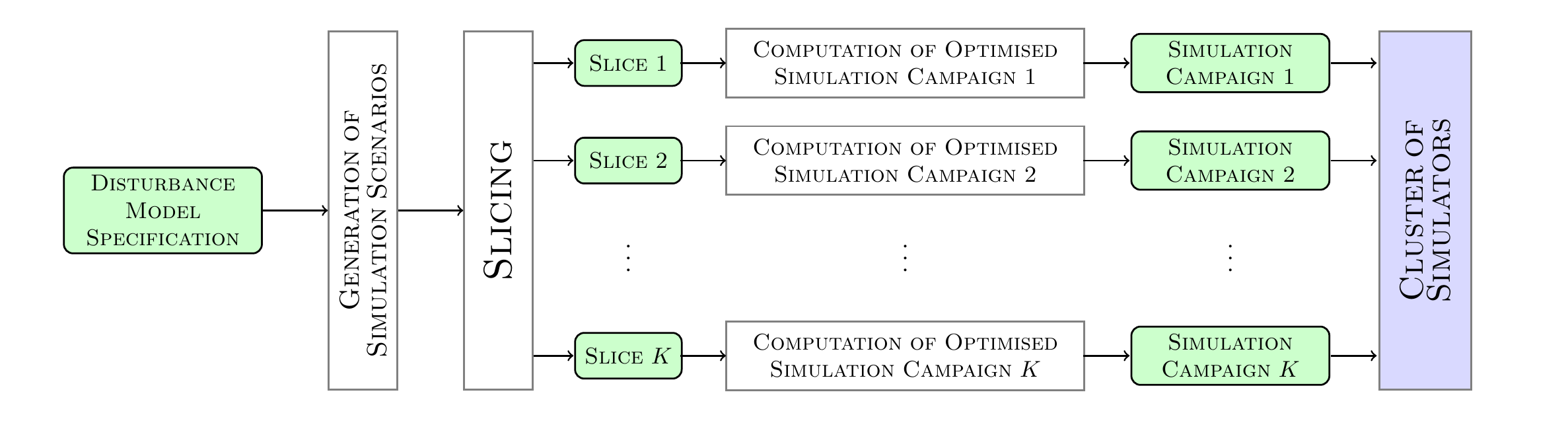}
	\caption{SLFV via Parallel Model Checking Driven Simulation\label{fig:intro:sylvaas}}
\end{center}
\end{figure}
\end{landscape}

The approach shown in Figure~\ref{fig:intro:sylvaas} is the so-called Parallel Model Checking Driven Simulation (MCDS) method, which performs SLFV by using a black box approach.
This is particularly useful to meet the need to exhaustively analyse the CPS behaviour
in all the relevant scenarios that may arise within its
operational environment.

This takes place during an initial offline phase where all such scenarios are generated.
This upfront generation enables the resulting scenarios to be
split into multiple chunks and then simulated in parallel~\cite{CAV2013}\cite{PDP2014}.
Hence, the MCDS method accelerates the entire SLFV process by involving multiple
computing resources which work simultaneously.

Furthermore, the generated scenarios are used to create optimised simulation
campaigns with the aim of avoiding redundant simulation.
Basically, these campaigns remove this redundancy by simulating duplicate scenarios only once. As a result, SBV is carried out more efficiently by reducing costs and time for simulation activities.

\clearpage
This optimisation is achieved by exploiting the ability of cutting-edge simulation
technology (\textit{e.g.}, Simulink\footnote{\url{https://www.mathworks.com/products/simulink.html}})
to store the intermediate states of the simulation\footnote{\textsc{Explanation of the store capability.} If the developer clicks \textsc{Start} in the Simulink toolbar, then the simulation begins. After a few seconds, if the developer clicks \textsc{Pause} in the same toolbar, then the simulation stops. At the same time, the simulator transparently stores the reached state of the simulation in the Simulink workspace. If the developer clicks \textsc{Start} again, then the simulator transparently loads the stored simulation state, and then starts the simulation from the exact simulation state that was reached when the developer clicked \textsc{Pause}. Fortunately, these actions can be performed programmatically or via command-line, without the need to open the Simulink GUI.}.
These intermediate states are then re-used as the starting point 
to simulate future scenarios, without the need to start from scratch~\cite{CAV2013}.

Experimental results in~\cite{PDP2014} show that by using 64 machines with an 8-core processor each, this parallel MCDS approach can complete the SLFV activity in about 27 hours whereas a sequential approach would require more than 200 days.

On top of that, the following benefits can also be exploited from
the MCDS method. For example, it can provide online information
about the verification coverage, and the Omission Probability (OP) \cite{DSD2014}\cite{Micro2016}.
Namely, it can indicate both the number of scenarios simulated so far,
and the probability of finding a yet-to-simulate scenario that
falsifies a given safety property.

The approach to SLFV that is closest to the MCDS method is presented in~\cite{Cav2013-6}, where the capability to call external C functions of the model checker CMurphi~\cite{DellaPenna2004} is exploited in a black box fashion to drive the ESA satellite simulator SIMSAT\footnote{\url{http://www.esa-tec.eu/space-technologies/from-space/real-time-simulation-infrastructure-simsat/}} in order to verify satellite operational procedures. Also in~\cite{Cav2013-16}, the analogous capability of the model checker SPIN\footnote{\url{http://spinroot.com/spin/whatispin.html}} is used to verify actual C code. Both these approaches differ from the MCDS method since they do not consider the optimisation of simulation campaigns.

\clearpage

\subsubsection*{Statistical and Monte Carlo model checking approaches}

Statistical model checking approaches, being basically black box, are also closely related to the MCDS method. For example, \cite{Cav2013-31} addresses SLFV of Simulink models and presents experimental results on the very same Simulink case studies that are used in this thesis. Monte Carlo model checking approaches (see, \textit{e.g.}, \cite{Cav2013-23}\cite{Cav2013-27}\cite{Cav2013-12}) are also related to this method as well.
The main differences between these two latter approaches and the MCDS method are the following: (i)~statistical approaches sample the space of admissible simulation scenarios, whereas MCDS addresses exhaustive SBV; (ii)~statistical approaches do not address optimisation of the simulation campaign, despite the fact that it makes exhaustive SBV more viable.

Formal verification of Simulink models has been widely investigated, examples are in~\cite{Cav2013-26}\cite{Cav2013-19}\cite{Cav2013-29}. These methods, however, focus on discrete time models (\textit{e.g.}, Simulink/Stateflow restricted to discrete time operators) with small domain variables. Therefore they are well suited to analyse critical subsystems, but cannot handle complex SLVF tasks (\textit{i.e.}, the case studies addressed in this thesis).

This is indeed the motivation for the development of statistical model checking methods as the one in~\cite{Cav2013-31} and for exhaustive SBV methods like the MCDS one. For example, in a Model Based Testing setting it has been widely considered the automatic generation of test cases from models (see, \textit{e.g.}, \cite{Cav2013-5}). In SBV settings, instead, automatic generation of simulation scenarios for Simulink has been investigated in~\cite{Cav2013-11}\cite{Cav2013-17}\cite{Cav2013-4}\cite{Cav2013-28}.
The main differences between these latter approaches and the MCDS method are the following. First, these approaches cannot be used in a black box setting since they generate simulation scenarios from Simulink/Stateflow models, whereas the MCDS method generates scenarios from a formal specification model of disturbances. Second, the above approaches are not exhaustive, whereas the MCDS method is.

Synergies between simulation and formal methods have been widely investigated in digital hardware verification. Examples are in~\cite{Cav2013-30}\cite{Cav2013-13}\cite{Cav2013-21}\cite{Cav2013-7} and citations thereof. The main differences between these latter examples and the MCDS method are: (i)~they focus on finite state systems, whereas the MCDS method focuses on infinite state systems (namely, hybrid systems); (ii)~they are white box (\textit{i.e.}, they require the availability of the CPS model source code) whereas the MCDS method is black box.

\clearpage
The idea of speeding up the SBV process by saving and restoring suitably selected visited states is also present in~\cite{Cav2013-7}. Parallel algorithms for explicit state exploration have been widely investigated. Examples are in~\cite{Cav2013-25}\cite{Cav2013-2}\cite{Cav2013-20}\cite{Cav2013-3}\cite{Cav2013-15}. The main difference between the MCDS method is that these latter ones focus on parallelising the state space exploration engine by devising techniques to minimise locking of the visited state hash table whereas the MCDS method leaves the state space exploration engine (\textit{i.e.}, the simulator)  unchanged, and uses an embarrassingly parallel (\textit{i.e.}, map and reduce like~\cite{Cav2013-8}) strategy that splits (map step) the set of simulation scenarios into equal size subsets to be simulated on different processors, and stops the SBV process as soon as one of these processors finds an error (reduce step).

The work in~\cite{Micpro2015-9} also presents an algorithm to estimate the coverage achieved using a SAT based bounded model checking approach. However, since scenarios are not selected uniformly at random, it does not provide any information about the OP, whereas the MCDS method does.

\subsubsection*{Random model checking, Coverage, and Omission Probability}

Random model checking is a formal verification approach closely related to the MCDS method. A random model checker provides, at any time during the verification process, an upperbound to the OP. Upon detection of an error, a random model checker stops and returns a counterexample. Random model checking algorithms have been investigated, \textit{e.g.}, in~\cite{Cav2013-12}\cite{Cav2013-27}\cite{Micpro2015-23}. The main differences with respect to the MCDS method are the following: (i)~all random model checkers generate simulation scenarios using a sort of Monte-Carlo based random walk. As a result, unlike the MCDS method, none of them is exhaustive (within a finite time horizon); (ii)~random model checkers (\textit{e.g.}, see~\cite{Cav2013-12}) assume the availability of a lower bound to the probability of selecting an error trace with a random-walk. Being exhaustive, the MCDS method does not make such an assumption.

The coverage yielded by random sampling a set of test cases has been studied by mapping it to the Coupon Collector's Problem (CCP) (see, \textit{e.g.},~\cite{Micpro2015-24}). In the CCP, elements are randomly extracted uniformly and with replacement from a finite set of $n$ test cases (\textit{i.e.,} simulation scenarios in the context of this thesis). Known results (see, \textit{e.g.},~\cite{Micpro2015-25}) tell us that the probability distribution of the number of test cases to be extracted in order to collect all $n$ elements has expected value of $\Theta$($n\log{}n$), and a small variance with known bounds. This allows the MCDS method to bound the OP during the SBV process.

Different from CCP based approaches, not only does the MCDS approach bound the OP, but it also grants the completion of the SBV task within just $n$ trials. This is made possible by the fact that simulation scenarios are completely generated upfront.

\subsection[Probabilistic Temporal Logic Falsification of CPSs]{Probabilistic Temporal Logic Falsification of\\Cyber-Physical Systems}

Another useful approach to speed up the verification process is
the so-called falsification method \cite{Abbas2013}.
Different from the approaches related to, and including, the MCDS method illustrated above, it uses techniques that are aimed at quickly identifying
scenarios that falsify the given properties.
In other words, these techniques search for operational scenarios where a given requirement specification is not met, hence the evaluation of the corresponding logical property is \textit{false}.
In this way, only a limited number of scenarios are considered,
thus less simulation is carried out.

In conclusion, both the aforementioned MCDS and the falsification methods use monitors
to discover potential simulation scenarios where the CPS being verified
violates a given specification (see, \textit{e.g.}, \cite{Cav2013-18}).
These monitors are typically developed manually in the language of the simulator
such as the Simulink and Matlab languages.
This is an error prone activity that can invalidate
the entire verification process,
because bugs in these monitors that are due to hand coding mistakes can lead to two significant errors. Firstly, they can result in false positives, \textit{i.e.}, simulation results of \textsc{Pass} which wrongly indicate that the system was able to work as expected under all the given operational scenarios. Secondly, they can result in false negatives, \textit{i.e.}, simulation results of \textsc{Fail} which wrongly points to a scenario were a given requirement was violated.

For this reason, ways to generate monitors directly from formal specifications
have recently been presented~\cite{Nickovic2007}\cite{Malter2008}\cite{Monitor3}\cite{Monitor4}\cite{Monitor5}.
Clearly, these automatically generated monitors are free of those bugs that are frequently introduced by hand coding.

\clearpage
\section{Motivations}

Despite the fact that SBV tools currently represent the main workhorse for SLFV, they
are unlikely to become more widely adopted,
unless two limiting factors are overcome.

First of all, the SLFV tool presented in~\cite{CAV2013} cannot optimise existing datasets, since they may
contain a huge number of collected scenarios. In fact,
no methods to obtain optimised simulation campaigns
from such datasets have yet been studied
in spite of the clear need to reduce the time for simulation activities while increasing the quality of these campaigns by removing redundant scenarios.

The second limitation that prevents wider use of current SLFV methods
is the lack of prior knowledge of the time needed to complete simulations.
The uncertainty surrounding the duration, makes scheduling SLFV activities problematic.
For example, underestimating the time needed for simulation activities could result in missed deadlines and costly project delays. Moreover, an overestimated time could result in meeting deadlines long beforehand, which means some resources allocated to the verification task would remain unused.

\section[Thesis Focus and Contributions]{Thesis Focus and Contributions to\\System-Level Formal Verification}

In this thesis, our main focus is on control systems modelled in Simulink\footnote{\url{https://www.mathworks.com/products/simulink.html}}, which is a widely used tool in the area of control engineering.

In order to overcome the aforementioned limitations in optimising existing datasets and estimating the time needed for SLFV tasks,
we devised and implemented the two following tools: (i)~a simulation campaign optimiser, and
(ii)~an execution time estimator for simulation campaigns.

\clearpage
\subsection[Simulation Campaign Optimiser]{Simulation Campaign Optimiser}

The optimiser tool is aimed at speeding up the SLFV process
by avoiding redundant simulation. To this end, we devised and implemented
a data-intensive algorithm to obtain optimised simulation campaigns from
existing datasets, that may contain a large quantity of collected scenarios.
These resulting campaigns are made up of sequences of simulator commands
that simulate common prefixes in the given dataset of scenarios only once.
These commands include \textit{save} a simulation state,
\textit{restore} a previously saved simulation state,
\textit{inject} a disturbance on the model, and
\textit{advance} the simulation of a given time length.

Results show that an optimised simulation campaign is at least 3 times
faster with respect to the non-optimised one~\cite{CAV2013}.
Furthermore, the presented tool can optimise 4~TB of scenarios
in 12 hours using just one core with 50~GB of RAM.

\subsection[Simulation Campaign Execution Time Estimator]{Execution Time Estimator for Simulation\\Campaigns}

The estimator tool is able to perform an accurate execution time estimate
using a partially machine-independent method.
In particular, it analyses the numerical integration steps
decided by the solver during the simulation.
From this preliminary analysis, it first chooses a small set of
simulator commands to simulate, and then it collects execution time samples
in order to train a prediction function.

Results show that this tool can predict the execution time to simulate
a given simulation campaign with an error below~10\%.

\clearpage
\subsection{Publications}

The following papers on our two main contributions presented in this thesis are currently being finalised for submission in collaboration with
\mbox{Vadim~Alimguzhin}, \mbox{Toni~Mancini}, \mbox{Federico~Mari}, \mbox{Igor~Melatti}, and \mbox{Enrico~Tronci}.

\begin{enumerate}
\item {\bf A Data-Intensive Optimiser Tool for Huge Datasets of Simulation Scenarios.} To be submitted to Automated Software Engineering.
\item {\bf Estimating the Execution Time for Simulation Campaigns with Applications to Simulink.} To be submitted to Transactions on Modeling and Computer Simulation.
\end{enumerate}

\clearpage
\section{Outline}

\begin{description}

\item[Chapter~\ref{chap:background}] illustrates background notions and
definitions that are used within the thesis.

\item[Chapter~\ref{chap:optimiser}] presents our optimiser tool.
In particular, it first shows the overall method we devised,
then it gives details of the algorithm we implemented,
finally it presents experimental results regarding the efficiency,
scalability, and effectiveness of the method.

\item[Chapter~\ref{chap:estimator}] introduces our estimator tool.
It gives a detailed description of the method we devised,
then it explains how we validated it, finally it presents the experimental results
of estimation accuracy.

\item[Chapter~\ref{chap:conclusions}] draws conclusions and
outlines future work.

\end{description}

%% file: chap02.tex
\chapter{Background}\label{chap:background}

\section{Notation}

In this thesis, we use $[n]$ to represent the initial segment $\lbrace 1, 2, \ldots , n\rbrace$ of natural numbers.
Often, we use $\mathbb{N}^+$ and $\mathbb{R}^+$ to denote the set of natural numbers and positive real numbers respectively.
Sometimes, the list \mbox{concatenation} operator is denoted by $\cup$ when we deal with sequences instead of sets.

\section{Definitions}

\subsection*{Disturbance}

\begin{definitionsec}[Disturbance]
A disturbance is a number $d \in \mathbb{N}^+ \cup \lbrace0\rbrace$. A disturbance identifies an exogenous event (\textit{e.g.} a fault) that we inject into the model being simulated. As an exception, when a disturbance $d$ is zero, we inject nothing on the model, thus a zero indicates what we call the \textit{non-disturbance} event.
\end{definitionsec}

\subsection*{Simulation Interval}

\begin{definitionsec}[Simulation Interval]
A simulation interval $\tau$ indicates a fixed amount of simulation seconds.
Note that the amount of simulation seconds is usually different from the elapsed time
the simulator takes to actually simulate them.
In fact, the elapsed time needed to simulate a given simulation interval $\tau$
may be higher or lower than $\tau$,
depending on the complexity of the simulation model.
\end{definitionsec}

\subsection*{Simulation Scenario}

\begin{definitionsec}[Simulation Scenario]
A simulation scenario (or simply scenario) is a finite sequence of disturbances $\delta = (d_1, d_2, \ldots, d_H)$ with length $|\delta|=H$.
These disturbances in $\delta$ are associated to $H$ consecutive simulation intervals that have fixed-length $\tau$.
In particular, we inject the model with each disturbance $d_i \in \delta$ at the
beginning of the $i$-th simulation interval. Hence, we simulate $\tau\cdot H$ simulation seconds starting from the initial simulator state, while injecting the model with disturbances at the beginning of each simulation interval.
\end{definitionsec}

\subsection*{Simulation Horizon}

\begin{definitionsec}[Simulation Horizon]
The simulation horizon (or simply horizon) is the length $H$ of a simulation scenario $\delta = (d_1, d_2, \ldots, d_H)$.
\end{definitionsec}

\subsection*{Dataset of Scenarios}

\begin{definitionsec}[Dataset of Scenarios]
A dataset of scenarios is a sequence $\Delta = (\delta_1, \delta_2, \ldots, \delta_N)$ that contains a number $|\Delta| = N$
of simulation scenarios. In particular, scenarios in $\Delta$ have horizon $H$. Furthermore, we denote each scenario $\delta_i \in \Delta$
by the sequence $\delta_i = (d^i_1, d^i_2, \ldots, d^i_H)$.
\end{definitionsec}

\subsection*{Simulation Campaign}

\begin{definitionsec}[Simulation Campaign]
Given a dataset of scenarios $\Delta$, a simulation campaign is a sequence of simulator commands that we use to 
simulate the input scenarios in $\Delta$. In particular, we use the five basic commands described in Table~\ref{fig:table:commands}.
\end{definitionsec}

\subsection*{Simulator Commands}

Table~\ref{fig:table:commands} shows the five basic simulator commands and their behaviour.

\renewcommand{\arraystretch}{1.2}
\begin{table}[H]
\begin{center}
\resizebox{1.1\textwidth}{!}{
\begin{tabular}[H]{|r|l|}
\hline
\hline
\textsc{Command} & \textsc{Behaviour}\\
\hline
\textit{inject}($d$) & {\em injects the simulation model with the given disturbance $d \in \mathbb{N}^+$}\\
\hline
\textit{run}($t$) & {\em simulates the model for $\tau\cdot t$ simulation seconds, with $t \in [H]$}\\
\hline
\textit{store}($x$) & {\em stores the current state of the simulator on a file named $x$}\\
\hline
\textit{load}($x$) & {\em loads the previously stored state file $x$ into the simulator}\\
\hline
\textit{free}($x$) & {\em removes the previously stored state file $x$}\\
\hline
\hline
\end{tabular}}
\caption{Simulator Commands\label{fig:table:commands}}
\end{center}
\end{table}

%% file: chap03.tex
\chapter{Optimisation of Huge Datasets of Simulation Scenarios}\label{chap:optimiser}

\section{Introduction}

\subsection{Framework}

The goal of Model Based System-Level Formal
Verification (SLFV) for Cyber-Physical Systems (CPSs) is
to prove that the system as a whole will behave as expected,
regardless of faulty events originating from the
environment it operates in.
The most widely used approach to SLFV
for complex CPSs is called Simulation Based Verification (SBV), and it is supported by
the increasing use of Model Based Design (MBD)
tools to develop CPSs.

Basically, SBV consists of simulating
both hardware and software components at
the same time, in order to analyse all
the relevant scenarios that the CPS
being verified must be able
to safely cope with.
Due to the risks associated with
errors in CPSs, it is extremely important
that SBV performs an exhaustive analysis of
all the relevant event sequences (\textit{i.e.}, scenarios) that
might lead to system failure.

SLFV uses formal languages to specify both the environment where the system
will operate, and the safety properties it is necessary to satisfy.
As a result of these specifications, SLFV can be performed using an assume-guarantee approach.
Namely, a successful SBV process can guarantee that
the CPS will behave as expected, assuming that 
all and only the relevant scenarios are formally specified.

Clearly, the earlier potential
errors are identified, the lower
the potential costs and effort
that are needed to fix them.
The fact that MBD tools enable
SBV activities to be carried out
in the early stages of development
life cycle is one of the main
reasons for their widespread
adoption.

Despite this significant advantage,
SBV is still an extremely time-consuming activity.
Simulation of complex CPSs can even
take as much as several months,
since it requires the analysis of
a considerable amount of
operational scenarios~\cite{PDP2014}.

Techniques aimed at speeding up
SBV have recently been explored~\cite{CAV2013}\cite{DSD2014}\cite{PDP2014}\cite{Micro2016}.
They basically exploit cutting-edge
simulation technology (\textit{e.g.},
Simulink\footnote{\url{https://www.mathworks.com/products/simulink.html}})
that allows for the storage of
intermediate states of
the simulation.
The aim is to reuse the intermediate
states as the basis for simulating
other future scenarios, instead of
having to start again from the
initial simulator state~\cite{CAV2013}.

This is achieved by performing an
upfront computation of optimised
simulation campaigns.
In particular, these are made of
sequences of simulator instructions
that are aimed at simulating
common prefixes between
scenarios only once, thus
removing redundant simulation.

Simulator instructions include
\textit{save} a simulation state,
\textit{restore} a saved simulation state,
\textit{inject} a disturbance on the model,
and \textit{advance} the simulation of
a given time length.

\subsection{Motivations}

Unfortunately, these optimised simulation
campaigns cannot currently be obtained
from existing datasets with a huge
number of collected scenarios.
The reason for this limitation is
that it is computationally expensive
to identify common prefixes in
large datasets of scenarios.

Consequently, scenarios used in~\cite{CAV2013}
are intentionally generated in a format that
makes it easy to spot common prefixes in
the resulting dataset.
They use a model checker to automatically
generate labelled lexicographically-ordered
sequences of disturbances, starting from
a formal specification model that
defines all such sequences.

\subsection{Contributions}

In the following, we show how
we overcome this limitation in
optimising existing datasets
with collected scenarios.
Namely, we present a data-intensive
optimiser tool to compute
optimised simulation campaigns
from these datasets.

This tool can efficiently perform
the optimisation of a large
quantity of scenarios that are
unable to fit entirely into
the main memory.
We accomplish this with a
data-intensive algorithm
that we describe in
detail in this chapter.

In particular, such an algorithm
involves a sequence of four distinct steps,
that we call {\em initial sorting},
{\em load labelling}, {\em store labelling},
and {\em final optimisation}.
We designed it in such a way as to
minimise both disk I/O latency and
the amount of intermediate output
data generated at each step,
similarly to that offered by current
big data computing frameworks such as
Apache Hadoop\footnote{\url{http://hadoop.apache.org/}}
and Apache Spark\footnote{\url{https://spark.apache.org/}}.

\section{Problem Formulation}

Starting from a dataset $\Delta$
with input scenarios, we generate the
corresponding optimised
simulation campaign.

The aim is to exploit the capability
of modern simulators to store and
restore intermediate states thus
removing the need to explore
common paths of scenarios
multiple times.

To be precise, let be given
a scenario $\delta = (d_1, \ldots, d_{H}) \in \Delta$.
The simulator runs under the
input $\delta$ starting from
its initial state.
Thus, any disturbance $d_i \in \delta$,
with $1 \leq i \leq H$, unequivocally
identifies the state reached by the
simulator immediately after the $i$-th
simulation interval of $\delta$.

Since many scenarios in $\Delta$
share common prefixes of simulation
intervals, many simulations which
explore the same states
multiple times can be
considered redundant.

For example, let be given two
scenarios $\delta = (\hat{d}_1,\ldots,\hat{d}_p, d_{p+1}, \ldots, d_{H})$,
and $\delta' = (\hat{d}_1, \ldots, \hat{d}_p, d'_{p+1}, \ldots, d'_{H})$.
They share the common prefix $(\hat{d}_1,\ldots,\hat{d}_p)$.
This prefix would normally be simulated
twice for both $\delta$ and $\delta'$.
This redundancy applies to
all the scenarios that share
common prefixes in $\Delta$.
In the following we show how we
avoid the simulation of common
prefixes multiple times
by exploiting the load/store
capabilities of modern simulators.

To simulate the first scenario
$\delta$, we: (i) run the simulator
with the input being the prefix
$(\hat{d}_1,\ldots,\hat{d}_p)$;
(ii) \textit{store} the simulator
state reached so far with a
label $\ell$; (iii) continue
the simulation of $\delta$ with
the input being the remaining
disturbances $(d_{p+1}, \ldots, d_{H})$.

To simulate the second scenario
$\delta'$, we: (i) \textit{load}
back the previously stored state $\ell$
thus avoiding the simulation of
the shared prefix $(\hat{d}_1,\ldots,\hat{d}_p)$;
(ii)~continue the simulation of
$\delta'$ with the input being
the remaining disturbances
$(d'_{p+1}, \ldots, d'_{H})$.

The difficulty in achieving this,
arises from the fact that input scenarios
in $\Delta$ are in a format that makes
the identification of such prefixes
computationally expensive.
In particular, input scenarios
in $\Delta$ are not lexicographically
ordered and do not contain labels
to identify common prefixes,
differently from the format
used in~\cite{CAV2013}.
Furthermore, $\Delta$ may contain
as many as millions of scenarios
that cannot be entirely
accommodated into the
available RAM.

\section{Methodology}

\subsection{Overall Method}

In the first step, we uniquely sort
the input dataset $\Delta$. The output
is a dataset $\mathcal{D}$ that contains
a number $|\mathcal{D}| = N$ of distinct
lexicographically ordered scenarios.
The resulting dataset helps us identify
common prefixes during the
following step.

In the second step, we compute a file
that contains a sequence
$\mathcal{L} = (\mathcal{L}_1, \mathcal{L}_2, \ldots, \mathcal{L}_N)$
with $N$ \textit{load labels}.
In particular, every label $\mathcal{L}_i$
corresponds to the label of the simulator
state that we load at the beginning
of the $i$-th scenario in
the resulting simulation campaign.
More precisely, $\mathcal{L}_i$
identifies the simulator state that
was previously stored immediately
after the simulation of the longest
shared prefix of disturbances between
the scenario $\delta_i \in \mathcal{D}$
and the very first scenario $\delta_j \in \mathcal{D}$,
with $j < i$, where such a
prefix appeared.

In the third step, we uniquely sort
the sequence $\mathcal{L}$. The output is
a file with a sequence $\mathcal{S}$ of
\textit{store labels}. This resulting sequence
helps us spot simulation states to
store during the final computation
of the simulation campaign
in the next step.

Finally, we use the input files
$\mathcal{D}$, $\mathcal{L}$, and $\mathcal{S}$
obtained so far in order to compute the final
optimised simulation campaign. Basically,
we compute a file that contains a
sequence $\mathcal{C}$ of simulator
commands that avoid the redundant
simulation of common paths of
scenarios in $\mathcal{D}$
multiple times.

\subsection{Steps of the Algorithm}

\subsubsection{Step 1. Initial Sorting}

In the first step, we compute
a dataset $\mathcal{D}$ of unique
lexicographically ordered disturbance
sequences, starting from the
input dataset $\Delta$, which contains
a multiset of unordered disturbance
sequences with fixed length $H$.

Since input scenarios in $\Delta$
do not fit entirely into the main memory,
an external sorting based approach~\cite{Knuth1998}\cite{Horowitz1983}
is used to order them.
Specifically, we first split these scenarios
into smaller chunks and sort them into
the main memory. Then, we iteratively merge
two sorted chunks at a time, until the last
two chunks are merged into the
resulting dataset $\mathcal{D}$. 

To this end, we allocate two buffers
of scenarios in RAM in order to minimise
the disk I/O latency. These buffers
also help to improve the performance
of the other steps where no sorting
is involved, namely Step 2 and
Step 4 described below.

\subsubsection{Step 2. Load Labelling}

In this step, we make use of the
following labelling strategy in order
to compute what we call the sequence
of \textit{load labels}
$\mathcal{L} = (\mathcal{L}_1, \mathcal{L}_2, \ldots, \mathcal{L}_N)$
from the dataset
$\mathcal{D} = (\mathcal{\delta}_1, \mathcal{\delta}_2, \ldots, \mathcal{\delta}_N)$
which we computed at Step 1.

Each label $\mathcal{L}_i$,
with $i \in [N]$, is associated
to the Longest Common Prefix (LCP)
of disturbances between the
$i$-th scenario $\delta_i$ and
all the previous scenarios
in $\mathcal{D}$.

Since scenarios in $\mathcal{D}$
are lexicographically ordered,
we can easily identify this LCP
just by comparing the $i$-th
disturbance sequence
$\delta_i = (d_1^i, \ldots, d_H^i)$
with the previous one
$\delta_{i-1} = (d_1^{i-1}, \ldots, d_H^{i-1})$,
with $i > 1$.

As a result, we obtain the
size $p(i)$ of this LCP that
we define as
$$p(i) := | \{ p \in [H] : \bigwedge\limits_{j = 1}^p d^i_j = d^{i-1}_j \} |.$$
$p(i)$ represents the index of
the rightmost disturbance of the
LCP between $\delta_i$ and
$\delta_{i-1}$.
Note that $p(i) = 0$ if no common
prefix exists, and also that
$p(i)$ is always less than $H$,
since $\mathcal{D}$ contains
no duplicate scenarios.

In order to compute each
label $\mathcal{L}_i$, we first
assign a label $\ell^i_j$,
with $j \in [H]$ to each
disturbance $d^i_j \in \delta_i$
in the following way:

$$\ell^i_j := \left\{\begin{array}{lr}
	\ell^{i-1}_j        & \text{if }j \leq p(i)\\
	(i-1)\cdot H + j    & \text{if }j > p(i)\\
	\end{array}\right.$$

Namely, the prefix of disturbances
$(d^i_1,d^i_2,\ldots,d^i_{p(i)})$
in $\delta_i$ share the same labels as
the shared prefix
$(d^{i-1}_1,d^{i-1}_2,\ldots,d^{i-1}_{p(i)})$
in the previous scenario $\delta_{i-1}$.
Instead, the remaining disturbances
$d^i_j$ in $\delta_i$, with
$j \in \lbrace p(i)+1,\ldots,H\rbrace$,
are strictly identified by
$(i-1)\cdot H+j$.

Finally, each label $\mathcal{L}_i$
corresponds to the exact label
associated to the rightmost disturbance
$d^i_{p(i)} \in \delta_i$ of the LCP,
hence:
$$\mathcal{L}_i := \left\{\begin{array}{lr}
	\ell^i_{p(i)} & \text{if }p(i) > 0\\
	0             & \text{if }p(i) = 0
	\end{array}\right.$$

Note that every label
$\mathcal{L}_i \in \mathcal{L}$,
with $\mathcal{L}_i\neq 0$, identifies
the simulator state reached
immediately after the
$(p(i)\text{ }\%\text{ }H)$-th
simulation interval of the
\mbox{$\lceil p(i)\text{ }/\text{ }H \rceil$-th}
scenario $\delta_{\lceil p(i)\text{ }/\text{ }H \rceil} \in \mathcal{D}$.
Also note that in order to compute
each label $\mathcal{L}_i$ there is
no need to entirely load $\mathcal{D}$
into the main memory.
In fact, all we need is to
compare the $i$-th scenario
with the previous one.

\subsubsection{Step 3. Store Labelling}

In this step, we compute a
unique lexicographically ordered
sequence $\mathcal{S}$ that contains what
we call the \textit{store labels},
starting from the sequence
$\mathcal{L}$ computed at Step 2.
Sorting is performed using the same
strategy we used at Step 1
to obtain $\mathcal{D}$.

The resulting sequence $\mathcal{S}$
contains the same labels in $\mathcal{L}$,
but each label in $\mathcal{S}$
appears only once.

We use this sequence of {\em store labels}
$\mathcal{S}$ during the computation
of the final simulation campaign.
In particular, $\mathcal{S}$ helps us
to spot all those simulator states
to store. Labels of {\em store}
instructions in the resulting
campaign appear in the same
order as they appear
in $\mathcal{S}$.

Note that the first label
in $\mathcal{S}$ is always $0$.
This is because there is at least
the first scenario in $\mathcal{D}$
which shares no common prefixes
with previous scenarios, thus
$\mathcal{L}_1 = 0$.
Also note that since the
label $0$ represents the
initial simulator state,
there is no need to store it.
In fact, we assume that this initial
state already exists before
the simulation begins.

\subsubsection{Step 4. Final Optimisation}

In this final step, we use the
files $\mathcal{D}$, $\mathcal{L}$,
and $\mathcal{S}$ obtained at
the previous steps in order to
compute the optimised
simulation campaign.

To this end, we build a sequence
$\mathcal{C} = \mathcal{C}_1\cup\text{ }\mathcal{C}_2\cup \ldots \cup\mathcal{C}_N$,
where each subsequence $\mathcal{C}_i$,
with $i \in [N]$, contains simulator
commands for the corresponding
scenario $\delta_i \in \mathcal{D}$.
Note that $\cup$ here denotes the list
concatenation operator, as
defined in Chapter~\ref{chap:background}.

We use the sequences of labels
$\mathcal{S}$ and $\mathcal{L}$ to add
commands \textit{store} and \textit{load}
respectively in the resulting
campaign, and are thus able to re-use
common simulation states
between scenarios.

In addition, we add some commands
\textit{free} in order to
remove previously stored
simulator states that
are no longer needed.

Algorithm~\ref{alg:compsimc} describes
the approach we use to perform
this final step.
We compute each subsequence
$\mathcal{C}_i$ of simulator
commands in the
following way.

First, we add the commands
\textit{free} (lines from 8 to 11)
in order to remove stored
simulator states that are
no longer needed.

Second, we add the command \textit{load}
(line 14) in order to restore the
previously stored simulator state
named $\mathcal{L}_i$.
Note that if $\mathcal{L}_i = 0$,
then the initial simulator
state is loaded and the
$i$-th scenario
$\delta_i \in \mathcal{D}$
is completely simulated
from the beginning.

Finally, we compute the commands
\textit{inject}, \textit{run}, and
\textit{store} (lines from 17 to 24).
To this end, we first build
a sequence of $k$ indexes
$(\mathcal{I}_1,\mathcal{I}_2,\ldots, \mathcal{I}_k)$
that has the following properties:
\begin{itemize}
	\item $\mathcal{I}_1$ is the index $p(i)$
		of the first time interval in $\delta_i$
		to simulate;
	\item $\mathcal{I}_k$ is always $H+1$ and
		we use it to add the last command \textit{run}
		(line~21), at the last iteration of
		the for-loop (line 17);
	\item All the other indexes
		$\mathcal{I}_j$, with $1 < j < k$,
		correspond to one of these two types
		of simulation intervals. The first
		are those intervals where we
		inject a disturbance $d^i_{\mathcal{I}_j}$
		(line 16), and the second
		are those ones where we add a
		command store (line 23) immediately
		after the simulator state
		reached by the previous
		command \textit{run} (line~21).
\end{itemize}

Note that there is no need to
entirely load $\mathcal{D}$, $\mathcal{L}$,
and $\mathcal{S}$ into the main memory.
In fact, in order to compute the
$i$-th sequence $\mathcal{C}_i$
of simulator commands,
all we need is the sequence
$(d^i_{p(i)+1}, d^i_{p(i)+2}, \ldots, d^i_{H}) \subseteq \delta_i$
(line~6), the label $\mathcal{L}_i$
(lines 5 and 14), and the first
$H-p(i)$ of the remaining
labels in $\mathcal{S}$
(lines 16 and 23).

\begin{algorithm}
\KwIn{Lex-ordered Dataset $\mathcal{D} = (\delta_1, \delta_2, \ldots, \delta_N)$.}
\KwIn{Sequence of \textit{Load Labels} $\mathcal{L} = (\mathcal{L}_1, \mathcal{L}_2, \ldots, \mathcal{L}_N)$.}
\KwIn{Lex-ordered Sequence of \textit{Store Labels} $\mathcal{S}$.}
\KwOut{Sequence of Simulator Commands $\mathcal{C}$.}

$\mathcal{C} \gets \varnothing$\;
$\mathcal{F} \gets (\mathcal{F}_0,\mathcal{F}_2,\ldots,\mathcal{F}_{H-1})$ such that $\mathcal{F}_i = 0$, with $i \in \lbrace 0,1,\ldots,H-1\rbrace$\;

\For{\textnormal{\textbf{each}} $i \gets 1, 2, \ldots, N$}{
	$\mathcal{C}_i \gets \varnothing$\;
	$p(i) \gets \mathcal{L}_i\text{ }\%\text{ }H$\;
	$(d^i_{p(i)+1},d^i_{p(i)+2},\ldots,d^i_{H}) \subseteq \delta_i \in \mathcal{D}$\;
	$\ell^i_j \gets (i-1)\cdot H + j$, with $j \in \lbrace p(i)+1,p(i)+2,\ldots,H\rbrace$\;

	\For{\textnormal{\textbf{each}} $j \gets p(i)+1,p(i)+2 \ldots, H-1$}{
		\If{$\mathcal{F}_j > 0$}{
			$\mathcal{C}_i \gets \mathcal{C}_i \cup \lbrace free(\mathcal{F}_j) \rbrace$\;
			$\mathcal{F}_j \gets 0$\;
		}
	}

	$\mathcal{C}_i \gets \mathcal{C}_i \cup \lbrace load(\mathcal{L}_i) \rbrace$\;
	$\mathcal{F}_{p(i)} \gets \mathcal{L}_i$\;

	$\mathcal{I} \gets (\mathcal{I}_1,\mathcal{I}_2,\ldots,\mathcal{I}_k)$ such that $\mathcal{I}_1 = p(i)+1$, $\mathcal{I}_k = H+1$, and $(d^i_{\mathcal{I}_j} \neq 0\text{ }\vee\text{ }\ell^i_{\mathcal{I}_j-1} \in \mathcal{S})$, with $\mathcal{I}_1 < \mathcal{I}_j < \mathcal{I}_k$\;

	\For{\textnormal{\textbf{each}} $j \gets 1, 2, \ldots, k-1$}{
		\If{$d^i_{\mathcal{I}_{j}} \neq 0$}{
			$\mathcal{C}_i \gets \mathcal{C}_i \cup \lbrace inject(d^i_{\mathcal{I}_{j}}) \rbrace$\;
		}
		$\mathcal{C}_i \gets \mathcal{C}_i \cup \lbrace run(\mathcal{I}_{j+1}-\mathcal{I}_j) \rbrace$\;
		\If{$\ell^i_{\mathcal{I}_{j+1}-1} \in \mathcal{S}$}{
			$\mathcal{C}_i \gets \mathcal{C}_i \cup \lbrace store(\ell^i_{\mathcal{I}_{j+1}-1}) \rbrace$\;
			$\mathcal{S} \gets \mathcal{S} \setminus \ell^i_{\mathcal{I}_{j+1}-1}$\;
		}
	}
	$\mathcal{C} \gets \mathcal{C} \cup \mathcal{C}_i$\;
}

\caption{Computation of the Optimised Simulation Campaign\label{alg:compsimc}}
\end{algorithm}

\clearpage
\subsection{Correctness of the Algorithm}

\subsubsection{Prefix-Tree of Scenarios}

Let's define $T_\mathcal{D} = (V, E)$
as the prefix-tree of scenarios
in $\mathcal{D}$.
In particular, $T_\mathcal{D}$
is a tree with height $H$, and
$V = \mathcal{S}$ $\cup$ $\lbrace 0 \rbrace$.
Namely, the set of vertexes
$V$ of $T_\mathcal{D}$ contains
the simulator state labels
in $\mathcal{S}$ including the
initial state label $0$,
which is the root of
$T_\mathcal{D}$.

The set of edges $E$ of $T_\mathcal{D}$
is composed in such a way that every path
$P^i = (0, e^i_1, v^i_1, e^i_2, v^i_2, \ldots, e^i_H, v^i_H)$ in $T_\mathcal{D}$
corresponds to the exact scenario
$\delta_i$ in $\mathcal{D}$ and vice versa,
with $i \in [N]$.
Clearly, every path $P^i$ starts
from the root $0$ and arrives
at one of the leaves
of $T_\mathcal{D}$.

In order to define
set $E$ more precisely, let's
define a function
$d : E \rightarrow \mathbb{N}^+ \cup\text{ }\lbrace 0 \rbrace$
that associates each edge $e \in E$
with a disturbance in $\mathbb{N}^+$
or $0$. Each edge in the path $P^i$
is an edge $e^i_j \in E$ such that
$d(e^i_j) = d^i_j \in \delta_i$,
with $j \in [H]$.

In conclusion, for all pairs
of scenarios $\delta_i, \delta_j$
in $\mathcal{D}$ that share a
common prefix $(d^i_1, d^i_2, \ldots, d^i_k)$
there exist the corresponding
paths $P^i, P^j$ in $T_\mathcal{D}$
that share their initial subpath
$(0, e^i_1, v^i_1, e^i_2, v^i_2, \ldots, e^i_k, v^i_k)$,
with $d(e^i_1) = d^i_1, d(e^i_2) = d^i_2, \ldots$,
and $d(e^i_k) = d^i_k$.

\subsubsection{Correctness of the Computed Subsequences}

The resulting simulation campaign
is a sequence $\mathcal{C}$ of simulator
commands.
We compute $\mathcal{C}$ by performing
an exploration of all the paths
$P^i$ in $T_\mathcal{D}$ using
an \textit{ordered} DFS-based
algorithm~\cite{Cormen2001}.
In particular, we visit all these
$P^i$ paths in the same order of
appearance as of the corresponding
scenarios $\delta_i$ in
$\mathcal{D}$.

During this exploration over
$T_\mathcal{D}$, we compute the
resulting simulation campaign
as follows.
For each scenario $\delta_i \in \mathcal{D}$,
we populate a subsequence
$\mathcal{C}_i$ with the
following simulator commands.
\begin{itemize}
	\item Command $load(\mathcal{L}_i)$
		(line 14). This command loads the
		previously stored simulator state
		$\mathcal{L}_i$. Note that
		$\mathcal{L}_i$ is a vertex
		of $T_\mathcal{D}$ which is	at
		height \mbox{$p(i) = \mathcal{L}_i\text{ }\%\text{ }H$}.

	\item Commands $free(\mathcal{F}_j)$,
		for some $j \in \lbrace p(i)+1, p(i)+2, \ldots, H-1\rbrace$
		(lines from 8 to 11).
		These commands remove all the
		previously stored simulator states
		that are no longer needed
		in the rest of the simulation campaign.
		In fact, these states are
		vertexes of $T_\mathcal{D}$ which are at
		a height which is greater than $p(i)$.
		As a result, all the paths that
		belong to subtrees of $T_\mathcal{D}$
		rooted at these vertexes are
		completely explored by the
		DFS at the exact moment
		when we append the command
		$load(\mathcal{L}_i)$ to
		$\mathcal{C}_i$ (line 14).

	\item Commands $inject(d^i_{\mathcal{I}_j})$,
		for some $j \in [k-1]$ (lines from 17 to 19).
		These commands inject the simulation
		model with each disturbance
		in the sequence
		$(d_{p(i)+1}, \ldots, d_H) \subseteq \delta_i$
		(line 6).

	\item Commands $run(\mathcal{I}_{j+1} - \mathcal{I}_j)$,
		for each $j \in [k-1]$ (line 21).
		These commands advance the
		simulator state by a number
		$\mathcal{I}_{j+1} - \mathcal{I}_j$ of
		consecutive simulation intervals
		of $\tau$ seconds each.
		Note that the argument $t$ of
		each $run(t)$ command (\textit{i.e.}, the
		number of simulation intervals)
		is decided in the sequence
		$\mathcal{I}$ (line 16).
		In particular, we compute this
		sequence $\mathcal{I}$ in such a way that
		between any two successive run
		commands in $\mathcal{C}_i$,
		there is either an inject
		(line 19) or a store
		command (line 23).

	\item Commands $store(\ell^i_{\mathcal{I}_{j+1}-1})$,
		for some $j \in [k-1]$ (line 23).
		These commands save the current
		simulator state on disk. Note that,
		for each index $\mathcal{I}_{j+1} \in \mathcal{I}$
		such that $\ell^i_{{\mathcal{I}_{j+1}}-1} \in \mathcal{S}$,
		we append $store(\ell^i_{{I_{j+1}}-1})$
		to save the simulator state reached
		immediately after the previous
		command $run(\mathcal{I}_{j+1} - \mathcal{I}_j)$
		(line 21).
\end{itemize}

\section{Experimental Results}\label{sec:results-optimiser}

In order to evaluate both
the efficiency and scalability
of our data-intensive
optimisation method, we
proceed as follows.

First, we optimise multiple
input datasets with different sizes
with the aim of analysing
the execution time of the
optimiser tool for each
input dataset.
In particular, we analyse
both the overall execution
time for optimisation
and the execution time of
the single steps of
the algorithm.

Finally, we analyse the
effectiveness of the method by
showing how much redundant simulation
can be removed in relation to
the input datasets we use.
Namely, we indicate how many
common prefixes are reused
in the final simulation campaign.

\subsection{Computational Infrastructure}

To run our experiments,
we use one single node in a 516-node
Linux cluster with 2500~TB of
distributed storage capacity.
This node consists of 2 Intel
Haswell 2.40~GHz CPUs with
8~cores and 128~GB of RAM.

\subsection{Benchmark}

We perform our experiments using
a 10~TB dataset of scenarios that
we automatically generate with the
model checker CMurphi \cite{DellaPenna2004}
from a formal specification model
that defines all the admissible
disturbances for the model
\textit{Apollo Lunar Module}\footnote{\url{https://it.mathworks.com/help/simulink/examples/developing-the-apollo-lunar-module-digital-autopilot.html}}
from the Simulink distribution.
In particular, this dataset contains
around 8.5~billion scenarios
with horizon $H = 150$ and
$12$ different kinds of
disturbance.

Note that scenarios in this dataset
are neither lexicographically
ordered nor labelled, differently
from the format of input
scenarios in~\cite{CAV2013}.

\subsection{Optimisation Experiments}

We use this 10-TB dataset
to obtain 10 smaller input datasets
of increasing sizes, from 100~GB
up to 5~TB. For each of these
input datasets, we run our
optimiser tool to compute
the corresponding optimised
simulation campaign.

In order to minimise the
disk I/O latency in reading
input scenarios, we allocate
50~GB of buffers in RAM
for each optimisation
experiment.

\begin{figure}[H]
\begin{center}
	\includegraphics[width=1\linewidth]{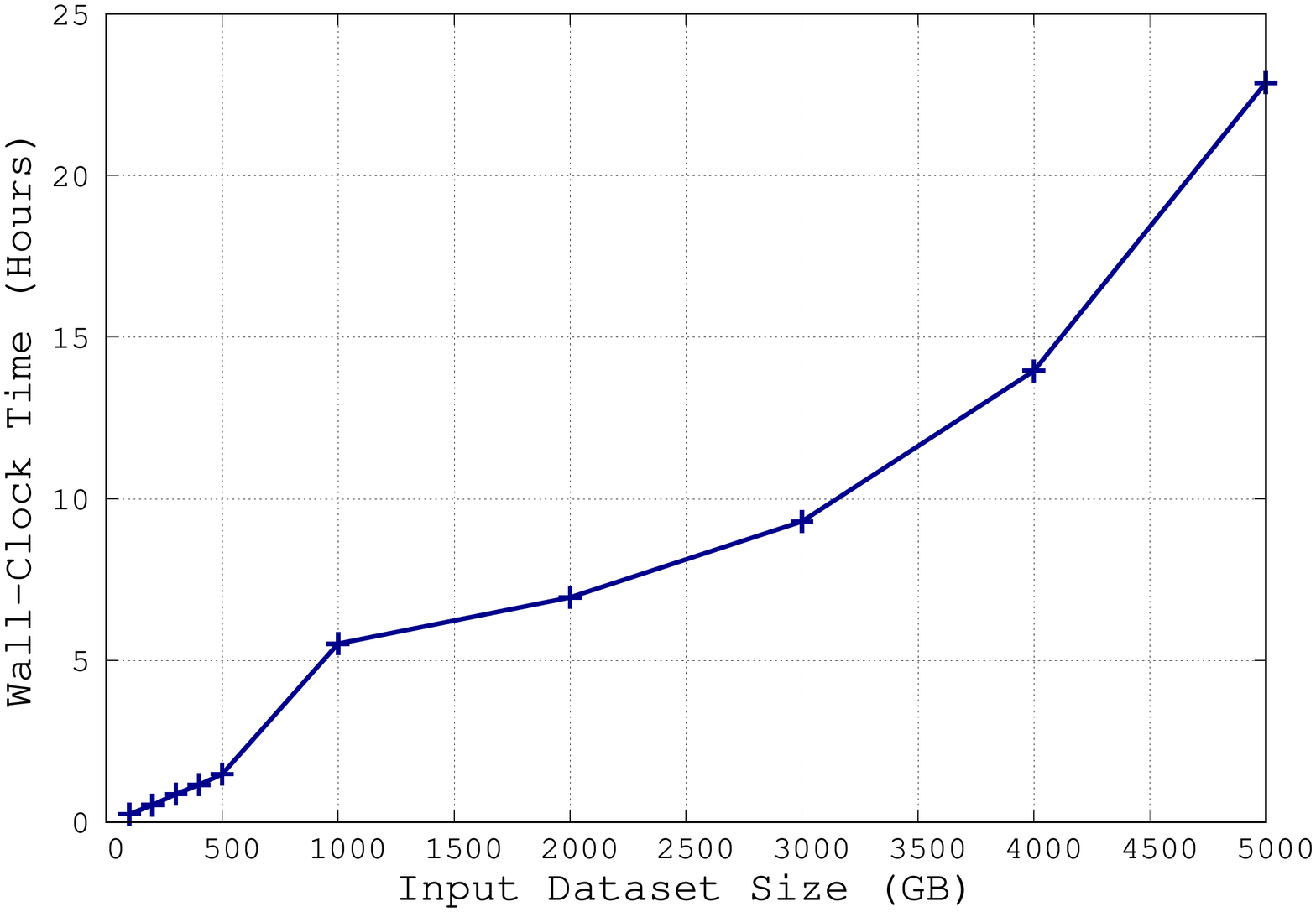}
	\caption{Overall Execution Time for Optimisation\label{fig:results:wallclock}}
	\end{center}
\end{figure}

\subsection{Execution Time Analysis}

Figure~\ref{fig:results:wallclock}
shows how the overall optimization time
grows in relation to the input size.
Table~\ref{fig:results:table} illustrates
the CPU time usage, which is useful to
understand how the I/O latency
contributes to the overall execution
time of our data-intensive algorithm.

Results from both
Figure~\ref{fig:results:wallclock}
and Table~\ref{fig:results:table}
indicate two significant outcomes.
Firstly, that our method is indeed
capable of optimising datasets of
disturbance traces that far exceed
the available amount of RAM.
Secondly, that the performance
decreases at a reasonable rate,
whereas input data grows at a significant rate
with respect to the amount
of RAM used for buffering.

\begin{figure}[H]
  \begin{center}
    \includegraphics[width=1\linewidth]{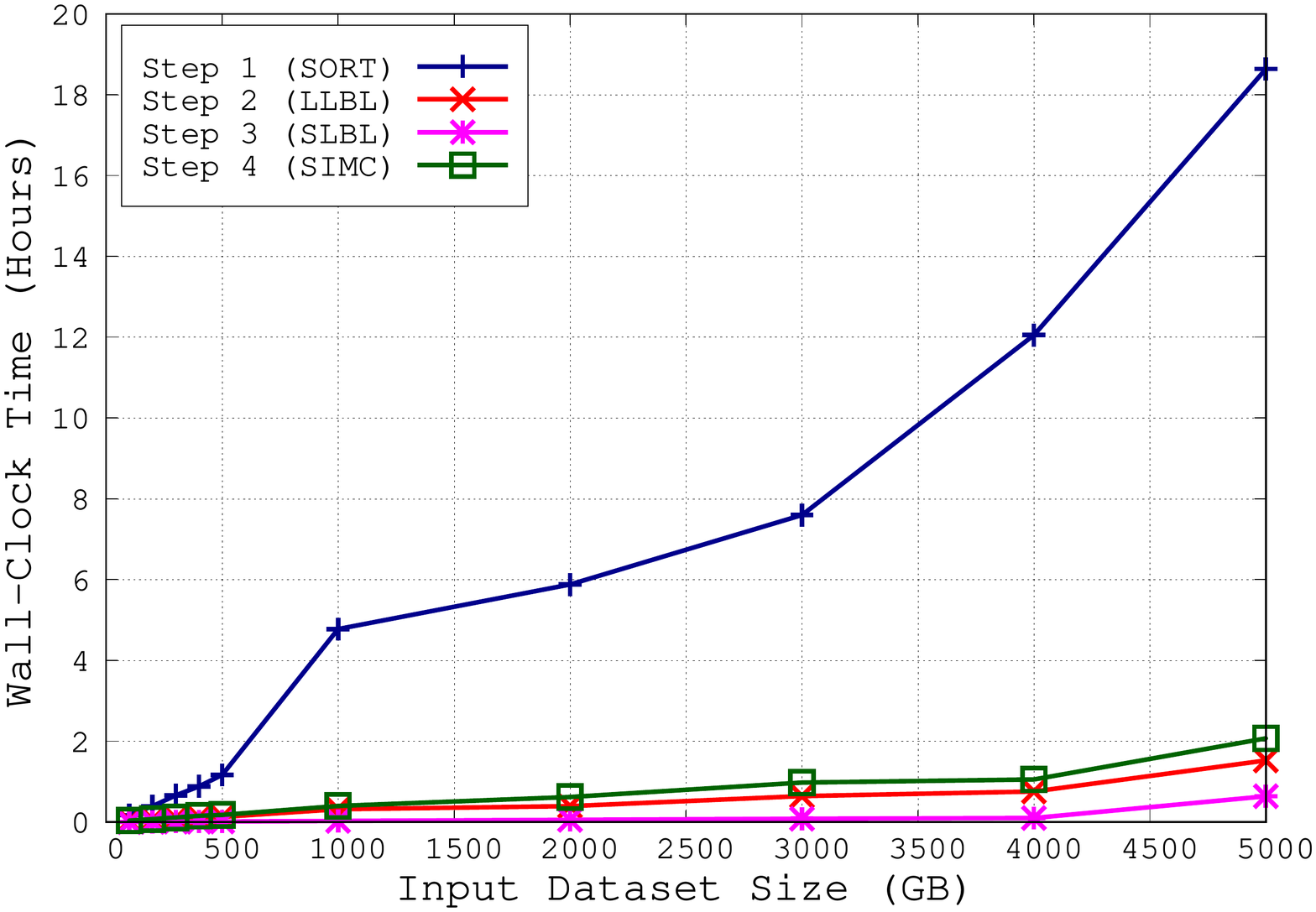}
    \caption{Execution Time for each Optimisation Step\label{fig:results:wallclock-steps}}
  \end{center}
\end{figure}

Figure~\ref{fig:results:wallclock-steps}
shows how the sorting step is significantly
more time-consuming than all the others.
In fact, it takes as long as 12~hours
to order the 4-TB input dataset,
6~times longer than the other
steps put together.

\begin{table}[H]
\begin{center}
\begin{tabular}{|r|r|r|r|r|}
\hline
\hline
Input (GB) & WC Time (sec) & Usr Time (sec) & Sys Time (sec) & CPU Perc. (\%)\\
\hline
100 & 869.96 & 256.97 & 208.19 & 53\\
\hline
200 & 1888.34 & 529.17 & 446.13 & 52\\
\hline
300 & 3123.07 & 807.58 & 728.63 & 49\\
\hline
400 & 4154.61 & 1082.31 & 1015.12 & 50\\
\hline
500 & 5330.37 & 1375.38 & 1346.83 & 51\\
\hline
1000 & 19860.45 & 2905.38 & 3056.02 & 30\\
\hline
2000 & 24998.79 & 6870.06 & 6922.10 & 55\\
\hline
3000 & 33479.43 & 9524.09 & 10505.79 & 60\\
\hline
4000 & 50248.18 & 12533.62 & 14587.52 & 54\\
\hline
5000 & 82325.26 & 17199.34 & 20858.58 & 46\\
\hline
\hline
\end{tabular}
\caption{Overall Execution Time and CPU usage for Optimisation\label{fig:results:table}}
\end{center}
\end{table}

\subsection{Optimisation Effectiveness Analysis}

In this section we evaluate the
effectiveness of our presented optimisation
method by analysing results
from our benchmark.
In particular, we show how much
redundant simulation can be removed
for each input dataset.
To this end, we analyse the number
of simulation intervals in both
the input scenarios and the
resulting simulation campaigns.

\begin{figure}[H]
  \begin{center}
    \includegraphics[width=1\linewidth]{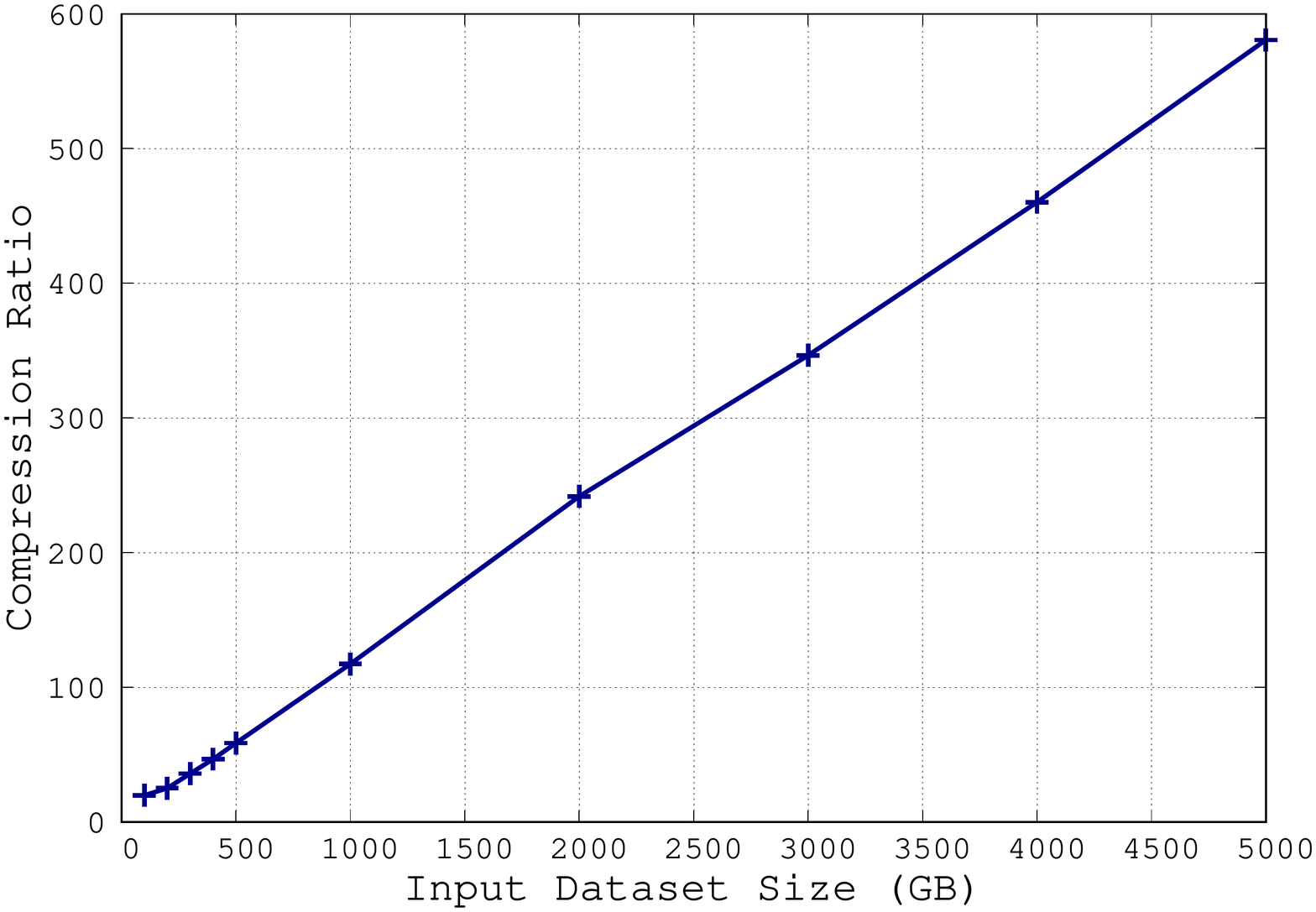}
    \caption{Compression Ratio per Input Dataset\label{fig:results:speedup}}
  \end{center}
\end{figure}

Figure~\ref{fig:results:speedup}
and Table~\ref{fig:results:table:compression}
show what we call the \textit{Compression Ratio}
between the input datasets and the
resulting optimised simulation
campaigns, which we define below.

First, let $SIM(\mathcal{D}):=N\cdot H$
be the number of simulation intervals in
a dataset $\mathcal{D}$ that contains
$|\mathcal{D}| = N$ input scenarios with
horizon $H$.
\mbox{Second}, let
$SIM(\mathcal{C}):=\sum_{run(t)\text{ }\in\text{ }\mathcal{C}} t$
be the number of simulation intervals
in the simulation campaign $\mathcal{C}$
obtained from $\mathcal{D}$, which is
the sum of the $t$-arguments of $run(t)$
commands in $\mathcal{C}$.
In conclusion, the Compression Ratio
between the input dataset and the
resulting optimised simulation campaign
is $SIM(\mathcal{D})/SIM(\mathcal{C})$.

\begin{table}[H]
\begin{center}
\begin{tabular}{|r|r|r|r|r|r|}
\hline
\hline
Input (GB) & \# of Scenarios & $SIM(\mathcal{D})$ & $SIM(\mathcal{C})$ & Compr. Ratio\\
\hline
100 & 83333333 & 12499999950 & 629628270 & 19.85\\
200 & 166666666  & 24999999900 & 994681374 & 25.13\\
300 & 250000000 & 37500000000 & 1043778055 & 35.93\\
400 & 333333333 & 49999999950 & 1070117781 & 46.72\\
500 & 416666666 & 62499999900 & 1067691791 & 58.54\\
1000 & 833330408 & 124999561200 & 1065522823 & 117.31\\
2000 & 1666663150 & 249999472500 & 1033963170 & 241.79\\
3000 & 2499995861 & 374999379150 & 1082233786 & 346.50\\
4000 & 3333326114 & 499998917100 & 1086509108 & 460.19\\
5000 & 4166658873 & 624998830950 & 1076531113 & 580.57\\
\hline
\hline
\end{tabular}
\caption{Compression Ratio per Input Dataset\label{fig:results:table:compression}}
\end{center}
\end{table}

\section{Related Work}

A method to compute optimised
simulation campaigns from datasets
of scenarios has already been
presented in \cite{CAV2013}.
The idea for this method was based
on the presence of labelled lexicographically
ordered datasets of input scenarios.
In particular, it exploited such
an input format in order to
efficiently build a Labels
Branching Tree (LBT) data
structure. The final simulation
campaign was then computed
from the LBT, which fitted
entirely in RAM.

As a result, the method
in~\cite{CAV2013} did not address
the computation of labels, and also did
not perform the ordering of
input scenarios.

In contrast, the method presented in
this chapter can efficiently compute
simulation campaigns from large
datasets of scenarios that
are neither labelled nor ordered,
and that cannot be fully accommodated into
the main memory.

\clearpage
\section{Conclusions}

In this chapter, we presented a
method to increase the performance
of System-Level Formal Verification (SLFV)
by computing highly optimised simulation
campaigns from existing datasets
that contain a huge number of
scenarios that do not fit entirely 
into the main memory.

For this purpose, we devised and
implemented a data-intensive algorithm
to efficiently perform the optimisation
of such datasets.

In fact, results show that a 4~TB dataset
of scenarios can be optimised in as little
as 12~hours using just one processor
with 50~GB of RAM.

%% file: chap04.tex
\chapter{Execution Time Estimation of Simulation Campaigns}\label{chap:estimator}

\section{Introduction}

\subsection{Framework}

Simulation Based Verification (SBV)
is currently the most widely used approach
to carry out System-Level Formal Verification (SLFV)
for complex Cyber-Physical Systems (CPSs).
In order to perform SBV, it is necessary to
execute simulation campaigns on the CPS model to be
verified.

Simulation campaigns consist of
sequences of simulator commands that are aimed at
reproducing all the relevant disturbance sequences (\textit{i.e.}, scenarios) originating
from the environment in which the CPS model
should operate safely.

Indeed, the goal of SLFV is to prove
that the system as a whole is able to
safely interact within this environment.
For this reason, simulation campaigns include
commands to \textit{inject} the model
with disturbances, which represent
exogenous events such as hardware
failures and system parameter
changes.

\clearpage
An example of a simulation campaign
could have the following sequence of instructions:
(i)~simulate the model for 3 seconds,
(ii)~inject the model with a disturbance,
(iii)~simulate the model for other 5 seconds, and
(iv)~check if the state of the model violates any given requirement.

Simulation campaigns are either
written by verification engineers
or automatically generated from a
formal specification model~\cite{CAV2013}.
In this latter case, simulation campaigns can
contain as many as $10^8$ of simulator commands,
which are stored in large binary
files of up to hundreds of gigabytes.

As a result of this specification model, simulation campaigns are generated so that SLFV can be performed using an assume-guarantee approach.
Namely, a successful SBV process can guarantee that
the CPS will behave as expected, assuming that 
all and only the relevant scenarios are formally specified.

\subsection{Motivations}

Simulation campaigns necessitate
a large number of
simulator instructions, consequently
the length of execution time required
to perform SBV tends to be considerable.
For example, as illustrated
in ~\cite{CAV2013} it takes as
long as 29 days to run a middle-sized
simulation campaign for the
Fuel Control System model\footnote{\url{https://www.mathworks.com/help/simulink/examples/modeling-a-fault-tolerant-fuel-control-system.html}}
with one 8-core machine (see also~\cite{DSD2014}\cite{PDP2014}\cite{Micro2016}).
This makes SBV an extremely
time-consuming process, and
therefore not very
cost effective.

This is mitigated by involving
multiple simulators working
simultaneously, in the form of
clusters that have hundreds
of nodes and thousands
of cores. For example, results in~\cite{PDP2014} show that by using 64 machines with an 8-core processor each, the SLFV activity can be completed in about 27 hours whereas a sequential approach would require more than 200 days.

However, there are still the following
obstacles to effective SLFV, that
involve the management of
computing resources.
In particular, clusters used to
perform SBV are either controlled
by commercial companies, thus
subjected to strict price policies,
or by community based research
institutions such as the
CINECA\footnote{\url{https://www.cineca.it/en}},
that offer free computation
hours for research activities.

\clearpage
Both these kinds of clusters
are typically constrained by:
\begin{enumerate}[label=(\roman*)]
	\item The number of cores per task that can be used;
	\item The amount of computation hours that can be used.
\end{enumerate}

As a result, it is necessary to either
manage the number of cores used to reduce
costs, or make the most of the limited number
of free computational hours available.
Furthermore, each verification phase
typically has a strict deadline to be
met in order to shorten
the time-to-market.

An accurate execution time estimate
can reveal if given deadlines can be met.
If not, mitigating actions can be
taken in good time. For example,
the process could be limited to a
subset of the simulation campaign.
Another option would be to buy
additional cores to execute the
entire simulation campaign faster.

On top of that, a prior knowledge of
the estimated execution time for a simulation
campaign helps in calculating the number of cores
and computation hours to be bought according
to the given time schedule and budget.
For example, it is helpful in deciding how
many slices to split the simulation campaign into
according to the given constraints.

In short, an accurate estimate of
execution time leads to better planning
of deadlines, and wiser budget
allocation for the required
computing resources.

\subsection{Contributions}

In order to overcome the lack of
tools to perform an effective estimation,
we show a method that uses just a small
number of execution time samples to
accurately train a prediction model.
In particular, we describe both how
we define our prediction model,
and how we choose simulator
instructions to collect
execution time samples.

This method brings two
significant advantages.

First, that only a very small
number of execution time samples
are required in order to train
the prediction model, thus removing
the need to simulate a
large simulation campaign.
Consequently the estimation process
is faster, and it is easier to make
decisions about the  number of
slices to split the campaign into
(\textit{i.e.}, how many machine
to use in parallel).

Secondly, the approach we use to
select which simulator commands to
sample is machine-independent.
Namely, decisions are made on the
basis of the simulation model and
the solver of the simulator.
This makes it possible to select
the simulator commands to train
the prediction model regardless
of the machine where the simulation
campaign will run. As a result,
we can select commands to sample
only once and reuse them to train
prediction models on each
different machine in
the cluster.

To the best of our knowledge,
there are no other applicable
methodologies in the literature.

\section{Background}

In this section we provide
some basic definitions that
will be used in the
rest of the chapter.

\subsection{Definitions}

\subsubsection{Simulation Model}

\begin{definitionsec}[Simulation Model]
A simulation model (or simply model)
$\mathcal{M}$ is a formal description
of the Cyber-Physical System (CPS) to
verify (\textit{e.g.}, the Fuel Control
System\footnote{\url{https://mathworks.com/help/simulink/examples/modeling-a-fault-tolerant-fuel-control-system.html}}).
In particular, the model
is written in the language of the
simulator being used.
\end{definitionsec}

\subsubsection{Simulator}

\begin{definitionsec}[Simulator]
A simulator $\sigma$ is a software tool
that is able to simulate the behaviour
of the given CPS model $\mathcal{M}$
to verify.
\end{definitionsec}

\subsubsection{Set of Disturbances}

\begin{definitionsec}[Set of Disturbances]
A set of disturbances $D_\mathcal{M}$
is a finite set of disturbances to inject
into the CPS model $\mathcal{M}$ during
the simulation.
Typically, each disturbance
$d \in D_\mathcal{M}$ consists in
modifying some system parameters
by the corresponding
simulator command $inject(d)$.
\end{definitionsec}

\subsubsection{Simulation Setting}

\begin{definitionsec}[Simulation Setting]
A simulation setting
$\mathcal{S} = (\sigma, \mathcal{M}, D_\mathcal{M}, \mu)$
identifies the simulator $\sigma$,
the model $\mathcal{M}$ to simulate,
the set of disturbances $D_\mathcal{M}$ to inject,
and the machine $\mu$ where the
simulator operates.
\end{definitionsec}

\subsection{Prediction Function}

Here we define prediction functions
and constants used to estimate the
execution time of each command in a
simulation campaign $\mathcal{C}$
for a given simulation
setting $\mathcal{S}$.

Function $run_\mathcal{S} : \mathbb{N}^+ \rightarrow \mathbb{R}^+$
represents the execution time estimate of
command $run(t)$ in the setting $\mathcal{S}$.
Namely, $run_\mathcal{S}(t)$ is the execution
time that the simulator takes to advance
the state of the model by $\tau\cdot t$
simulation seconds.

Constants
	$inject_\mathcal{S}$,
	$store_\mathcal{S}$,
	$load_\mathcal{S}$,
	$free_\mathcal{S} \in \mathbb{R}^+$ 
represent the execution time estimate of commands
	\textit{inject},
	\textit{store},
	\textit{load}, and
	\textit{free}, in the same setting $\mathcal{S}$.

Based on a simulation setting $\mathcal{S}$,
a simulation campaign $\mathcal{C}$,
and a command $c \in \mathcal{C}$,
we define as follows our prediction function
$\mathcal{P}_\mathcal{S}(c)$ to indicate the
execution time estimate for command
$c$ in the simulation setting $\mathcal{S}$.
$$\mathcal{P}_\mathcal{S}(c) = \left\{\begin{array}{ll}
	run_\mathcal{S}(t)		& c = run(t), t \in \mathbb{N}^+\\
	inject_\mathcal{S}		& c = inject(d), d \in D_\mathcal{M}\\
	store_\mathcal{S}		& c = store(s), s \in \mathbb{N}^+\\
	load_\mathcal{S}		& c = load(s), s \in \mathbb{N}^+\\
	free_\mathcal{S}		& c = free(s), s \in \mathbb{N}^+\\
	\end{array}\right.$$

\clearpage
\section{Problem Formulation}

Let be given a simulation setting
$\mathcal{S}$ and a simulation campaign
$\mathcal{C}$. The aim is to automatically
find an accurate estimation of the
execution time to simulate $\mathcal{C}$
in the simulation setting $\mathcal{S}$ by training
our prediction function $\mathcal{P}_\mathcal{S}$
in such a way that the resulting estimate
$\mathcal{P}^\mathcal{C}_\mathcal{S}$ is the sum
of the estimates for each simulator command $c$
in $\mathcal{C}$, \textit{i.e.},
$$\mathcal{P}^\mathcal{C}_\mathcal{S} = \sum_{c\text{ }\in\text{ }\mathcal{C}}\mathcal{P}_\mathcal{S}(c).$$

In the rest of the chapter we
show how we effectively train
$\mathcal{P}_\mathcal{S}$ and
how we validate its accuracy.

\section{Methodology}

Given how useful having an
execution time estimate is, it would be
tempting to compute it on the basis
of an existing simulation campaign.
Once collected, the execution time
samples would be used to train a
prediction model. However,
this na{\"i}ve approach is
risky as it could potentially
lead to either poor performances
or to inaccurate results for the
following reasons. First, the given
simulation campaign may require
hours of simulations.
Second, if a relatively small subset of
the given campaign is used, the
resulting collected samples could be
either insufficient or inappropriate to train
an accurate prediction model.

As an example, let us consider
only the command $run(t)$, which
is the most expensive from a computational point of view.
It would be natural to assume a linear
relationship between the execution
time $e_{run}(t)$ to actually
simulate it, and the number of 
simulation seconds $\tau\cdot t$ that are
required to advance. Namely, the estimated
execution time to simulate $run(t)$ would be
$e_{run}(t) := a\cdot t + b$.

To train such a simple prediction model,
a careful selection of $t$-values to sample
is needed, according to the following
phenomena we observed on
several Simulink models.

First, for small values of $t$ up
to a certain value $t_{min}$, the execution
time to simulate $run(t)$ remains reasonably
constant or it grows at a very low rate.
Second, for large values of $t$ from a
certain value $t_{MAX}$, the execution
time grows constantly and it is strictly
related to the number of numerical
integration steps performed by the
simulator\footnote{Simulators generally use Ordinary
Differential Equations (ODE) solvers that implement
state-of-the-art algorithms for numerical integration.
These algorithms are designed to reduce
simulation time while maintaining
high numerical accuracy by dynamically
choosing the integration steps
to perform (\url{https://www.mathworks.com/help/matlab/ordinary-differential-equations.html}).}.
Clearly, $t_{min}$ and $t_{MAX}$ vary depending
on the complexity of the model being simulated.

Note that a typical simulation
campaign includes a large number of $run(t)$
commands with both short and
long values of $t$.
These last observations together constitute
the main reason why the aforementioned
na{\"i}ve approach would lead to
inaccurate estimates, if execution
time samples are not collected properly.

\subsection{Overall Method}

In order to train our prediction function $\mathcal{P}_\mathcal{S}$, we use a small number of execution time samples that are sufficient to make an accurate execution time estimate of a given simulation campaign $\mathcal{C}$. To this end, we collect such samples by simulating an ad-hoc sequence $\mathcal{C}^*$ of simulator commands that we choose for this purpose.

More precisely, we proceed as follows.
First, we accurately choose simulator commands to put in $\mathcal{C}^*$ in order to compute a meaningful training set.
Second, we simulate $\mathcal{C}^*$ to collect execution time samples in our training set.
Finally, we use the resulting training set to compute parameters for our prediction function $\mathcal{P}_\mathcal{S}$.

In the following sections we show how we build $\mathcal{C}^*$ and more specifically which $t$-values
we choose to collect execution time samples of $run(t)$ commands.

\clearpage
\subsection{Training Phase}

\begin{landscape}
\begin{figure}
	\center
	\includegraphics[width=0.75\linewidth]{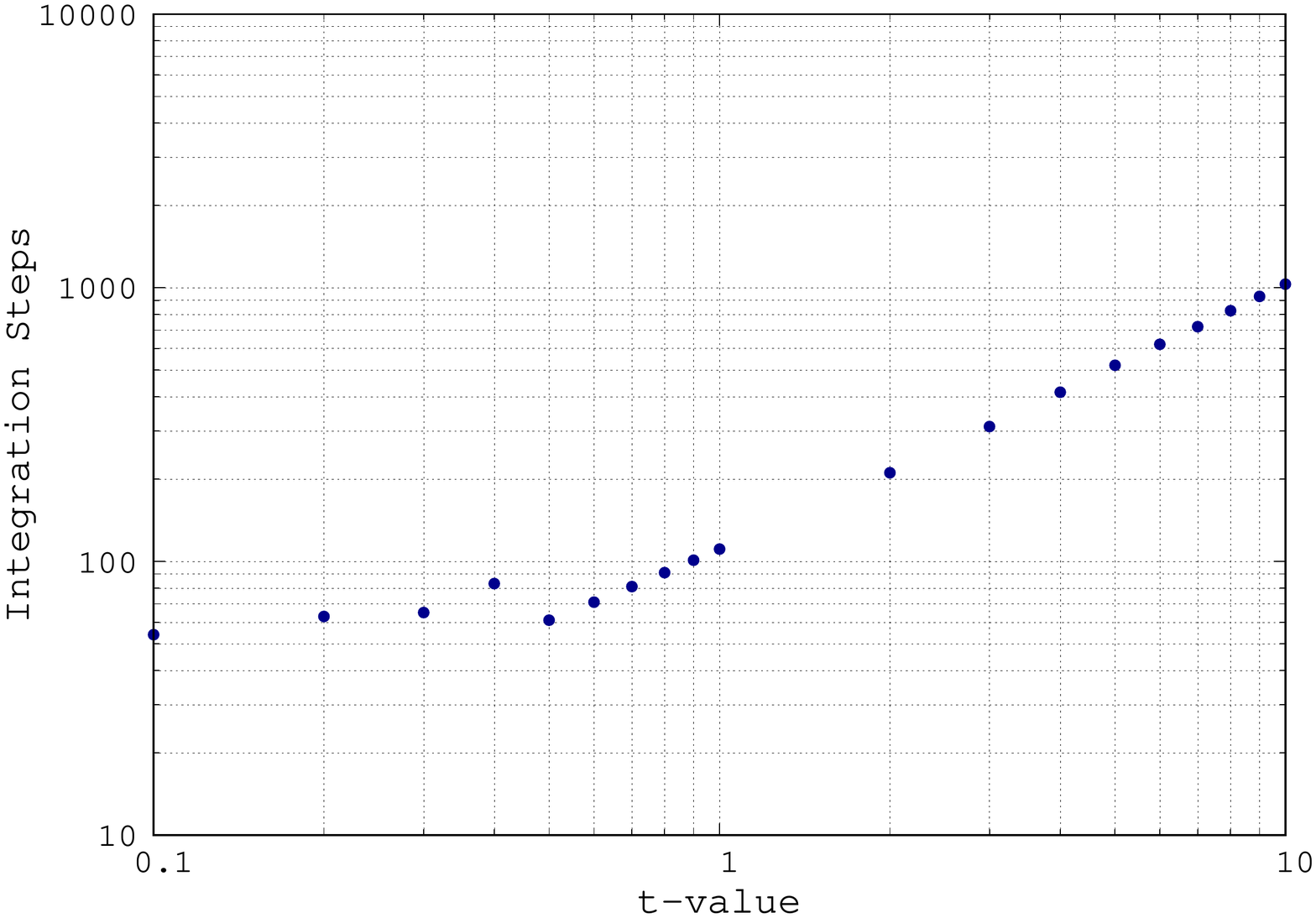}
	\caption{Integration Steps to Simulate Command $run(t)$\label{fig:methodology:steps}}
\end{figure}
\end{landscape}

\begin{landscape}
\begin{figure}
	\center
	\includegraphics[width=0.75\linewidth]{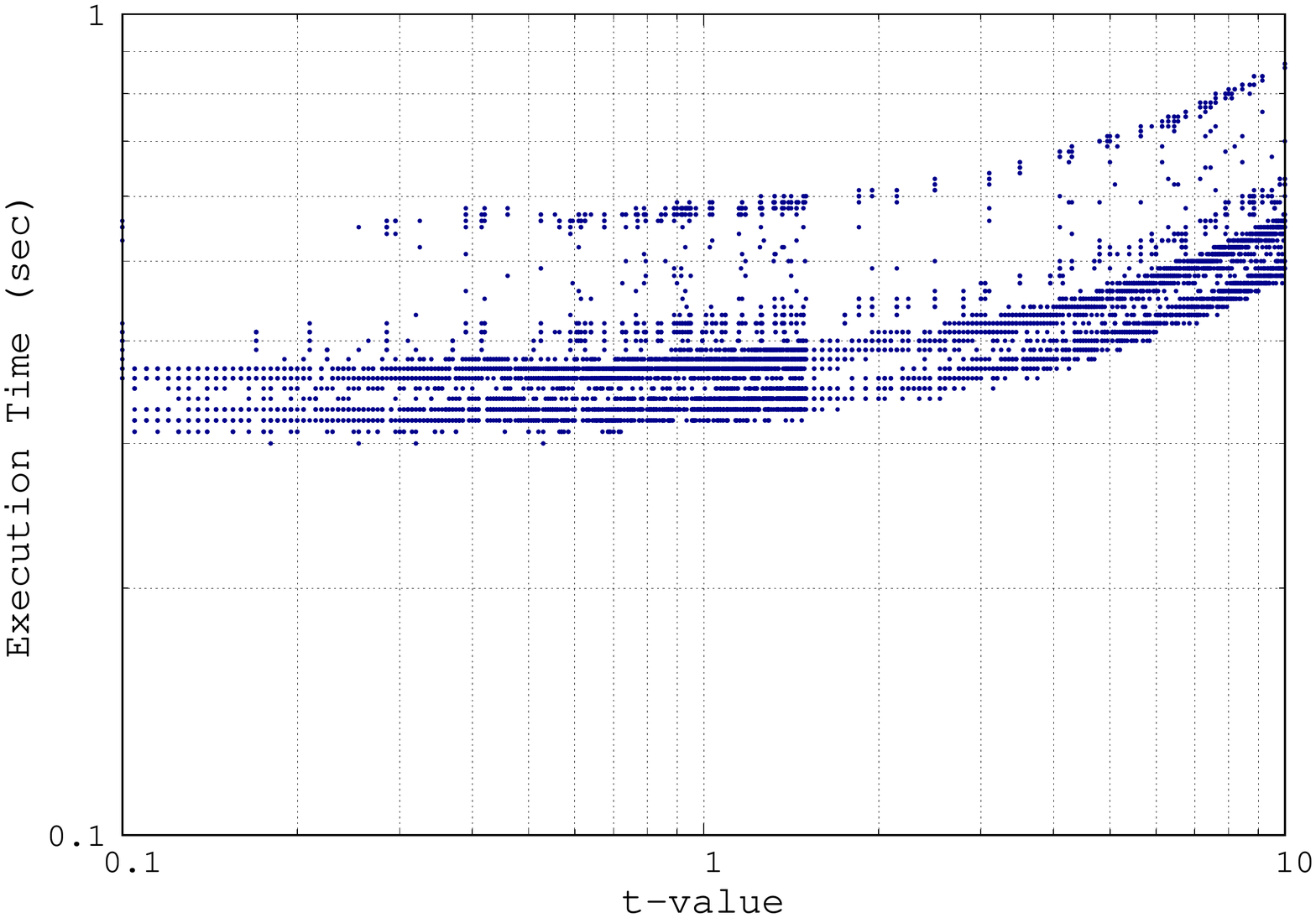}
	\caption{Execution Time to Simulate Command $run(t)$\label{fig:methodology:run}}
\end{figure}
\end{landscape}

When working with Simulink, if we simulate a command $run(t)$ multiple times with the same $t$-value on a model $\mathcal{M}$, 
then the solver always chooses the same numerical integration steps, hence the number of such steps remains unchanged (Figure~\ref{fig:methodology:steps}).
Clearly, this stability is not reflected in the resulting execution time when we simulate the same command $run(t)$ multiple times with the same $t$-values (Figure~\ref{fig:methodology:run}).

Note that Figure~\ref{fig:methodology:steps} and Figure~\ref{fig:methodology:run} are similar
in that they are almost constant for $t \leq 1$ and they linearly grow for $t > 1$.
This similarity can be explained by the fact that the execution time the simulator needs
is strictly related to the numerical integration steps it performs.
Also note that the breakpoint at $1$ in these figures is strictly linked to
the model $\mathcal{M}$ being simulated, which is the Fuel Control System in this case.

Based on the above considerations, we define our prediction function $run_\mathcal{S}$ that estimates
the execution time needed to simulate $run(t)$ as the following 2-step Piecewise Linear Function (PWLF).
$$run_\mathcal{S}(t) := \left\{\begin{array}{lr}
	\alpha                        & t \leq \gamma\\
	\beta\cdot(t-\gamma) + \alpha & t \geq \gamma\\
	\end{array}\right.$$

In order to train this function, we first collect execution time samples 
of $run(t)$ commands, and then we use these samples to find the intercept $\alpha$, the slope $\beta$, and
the breakpoint $\gamma$ of $run_\mathcal{S}$.

In the following section we illustrate how we choose these $t$-values by analysing the integration steps that 
the simulator performs for each $t$-value.

\subsubsection{Set of $t$-values $T_\mathcal{M}$}

The aim is to select a set $T_\mathcal{M}$ with the $t$-values of $run(t)$ commands to simulate in order to collect meaningful training data for our prediction function $run_\mathcal{S}$. In particular, we need to collect execution time samples of $run(t)$ commands for either short or long $t$-values.

To this end, we first find two values $t^{min}_\mathcal{M}$ and $t^{MAX}_\mathcal{M}$, that correspond to the minimum and the maximum $t$-values in $T_\mathcal{M}$ respectively.

Let define $N_\mathcal{M}(t)$ as the number of integraton steps that the simulator performs to simulate $run(t)$ on the model $\mathcal{M}$.
Our method to find $t^{min}_\mathcal{M}$ and $t^{MAX}_\mathcal{M}$ exploits these two key observations: (i) for small $t$-values, the execution time to simulate $run(t)$ remains relatively constant, \textit{i.e.}, $run_\mathcal{S}(t) \simeq c_1$, for $t < t^{min}_\mathcal{M}$, and (ii) for large $t$-values,
it grows constantly and it is strictly related to the number $N_\mathcal{M}(t)$ of integration steps,
\textit{i.e.}, $run_\mathcal{S}(t) \simeq N_\mathcal{M}(t)\cdot c_2$, for $t > t^{MAX}_\mathcal{M}$.

To find $t^{min}_\mathcal{M}$, we use the Algorithm~\ref{algorithm:tmin}, which works in the following way. Basically, we search for the
largest interval $(0, t^{min}_\mathcal{M})$ of all the $t$-values for which the number $N_\mathcal{M}(t)$ of steps performed to
simulate $run(t)$ remains constant. Note that for $t$-values that are close to zero, the simulator performs the same number $N_\mathcal{M}(t)$ of integration steps.
For these reasons, in order to train $run_\mathcal{S}(t)$ it is sufficient to choose $t^{min}_\mathcal{M}$ as the greatest $t$-value such that
$N_\mathcal{M}(t)$ remains constant for every $t'$ less than or equal to $t^{min}_\mathcal{M}$, \textit{i.e.},
$$t^{min}_\mathcal{M} = \max \{t \;|\; (\forall t' \leq t)(N_\mathcal{M}(t) = N_\mathcal{M}(t')) \}.$$
Once we have selected the lower bound $t^{min}_\mathcal{M}$ of $T_\mathcal{M}$, we look for the smallest upper bound $t^{MAX}_\mathcal{M}$
in order to make the final training set as small as possible, thus minimising the training effort.

To find $t^{MAX}_\mathcal{M}$, we use the Algorithm~\ref{algorithm:tmax}.
As mentioned,
the function $N_\mathcal{M}(t)$ is relatively constant for small $t$-values, while it linearly grows for the larger ones.
For this reason, the value of $t^{MAX}_\mathcal{M}$, is the smallest $t$ such that the linear regression on the
collected points in $(t^{min}_\mathcal{M}, t]$ has an error lower than a given bound $\varepsilon$.

\clearpage
\begin{algorithm}
\KwIn{Model $\mathcal{M}$.}
\KwOut{$t^{min}_\mathcal{M} \in \mathbb{R}^+$.}
$N_0 \gets N_\mathcal{M}(10^0)$\;
\For{\text{\normalfont \textbf{each}} $i \gets -1, -2, -3, \ldots$}{
  $N_i \gets N_\mathcal{M}(10^{i}))$\;
  \If{$N_i = N_{i+1}$}{
    \textbf{exit for}\;
  }
}
$t^{min}_\mathcal{M} \gets 10^{i+1}$\;
\caption{Find $t^{min}_\mathcal{M}$.}
\label{algorithm:tmin}
\end{algorithm}

\begin{algorithm}
\KwIn{Model $\mathcal{M}$; $t^{min}_\mathcal{M} \in \mathbb{R}^+$; $\varepsilon$.}
\KwOut{$t^{MAX}_\mathcal{M} \in \mathbb{R}^+$.}
\For{\text{\normalfont \textbf{each}}  $i \gets 1, 2, 3, \ldots$}{
  $t_i \gets t^{min}_\mathcal{M}\cdot 10^{i-1}$\;
  $N_i \gets N_\mathcal{M}(t_i)$\;
  \If{$i \ge 3$}{
    $\tilde{N}_i \gets \alpha + \beta \cdot t_i$, where $\alpha$ and $\beta$ are linear regression coefficients
    	computed from a set of samples $(t_j, N_j)$ collected at previous iterations, with $j = 1, 2, \ldots, i-1$\;
    \If{$|\frac{N_i}{\tilde{N}_i}-1| < \varepsilon$}{
    	\textbf{exit for}\;
    }  
  } 
}
$t^{MAX}_\mathcal{M} \gets t_i$\;
\caption{Find $t^{MAX}_\mathcal{M}$.}
\label{algorithm:tmax}
\end{algorithm}

\clearpage
From $t^{min}_\mathcal{M}$ and $t^{MAX}_\mathcal{M}$, we can finally define our set of $t$-values as the following.
Let $a$ and $b$ be such that $t^{min}_\mathcal{M} = 10^a$ and $t^{MAX}_\mathcal{M} = 10^b$ respectively.
$$T_\mathcal{M} := \bigcup^{b-1}_{i = a}\bigcup^{10}_{j = 1} \lbrace 10^i\cdot j \rbrace.$$
Namely, we put in $T_\mathcal{M}$ $10$ equidistant $t$-values from all the subranges $[10^i, 10^{i+1}]$ in the range $[t^{min}_\mathcal{M}, t^{MAX}_\mathcal{M}]$, with $i = a, a+1, \ldots, b-1$.

Table~\ref{fig:table:samplepoints} shows the values $t^{min}_\mathcal{M}$ and $t^{MAX}_\mathcal{M}$ found by the Algorithm~\ref{algorithm:tmin} and the Algorithm~\ref{algorithm:tmax} including the error  $\varepsilon$ used by the Algorithm~\ref{algorithm:tmax}, and the resulting total number $| T_\mathcal{M} |$ of $t$-values to sample, for each Simulink model.

Note that the computation of $T_\mathcal{M}$ is machine-independent. In fact, the information we use to find $t^{min}_\mathcal{M}$ and $t^{MAX}_\mathcal{M}$ depends only on the given model $\mathcal{M}$ and the simulator solver.

\begin{table}
\begin{center}
\begin{tabular}[H]{|l|l|l|l|l|}
\hline
\hline
Model $\mathcal{M}$ &  $t^{min}_\mathcal{M}$ & $t^{MAX}_\mathcal{M}$ & $\varepsilon$ & $| T_\mathcal{M} |$\\
\hline
\hline
\texttt{sldemo\_fuelsys.mdl} & $10^{-2}$ & $10^3$ & $10^{-2}$ & $46$\\
\hline
\texttt{aero\_dap3dof.slx} & $10^{-4}$ & $10^2$ & $10^{-2}$ & $55$\\
\hline
\texttt{penddemo.slx} & $10^{-6}$ & $10^2$ & $10^{-2}$ & $73$\\
\hline
\texttt{sldemo\_boiler.slx} & $10^{1}$ & $10^5$ & $10^{-2}$ & $37$\\
\hline
\texttt{sldemo\_engine.slx} & $10^{-5}$ & $10^4$ & $10^{-2}$ & $82$\\
\hline
\texttt{sldemo\_househeat.slx} & $10^{-1}$ & $10^4$ & $10^{-2}$ & $46$\\
\hline
\hline
\end{tabular}
\caption{Values of $t$ to Sample for Command $run(t)$\label{fig:table:samplepoints}}
\end{center}
\end{table}

\subsubsection{Simulation Campaign $\mathcal{C}^*$}

From the set $T_\mathcal{M}$ of $t$-values, we populate the sequence $\mathcal{C}^*$ of simulator commands.
In particular, this sequence is made up of commands $inject(d)$ and $run(t)$, for each $t \in T_\mathcal{M}$, and for each disturbance $d \in D_\mathcal{M}$.
On top of that, we populate $\mathcal{C}^*$ with commands \textit{load}, \textit{store}, and \textit{free} in a consistent way. Namely, commands $load(x)$ and $free(x)$ are preceded by the 
corresponding command $store(x)$.

Once our simulation campaign $\mathcal{C}^*$ is ready, we use it to populate our training set with execution time samples.
To this end, we simulate $\mathcal{C}^*$ multiple times in order to collect a meaningful number of training samples.

\subsection{Prediction Function}

\subsubsection{Command Run}

In order to train our prediction function $run_\mathcal{S}$, we proceed in the following way.
Our training set $S_\mathcal{M} := (S_1, S_2, \ldots, S_N)$ contains
those samples collected from simulating $run(t)$ commands in the simulation campaign $\mathcal{C}^*$. Namely, each $S_i = (t_i, \tau_i)$, with $i \in [N]$, is a pair that contains the measured execution time $\tau_i$ to simulate $run(t_i)$ on the model $\mathcal{M}$, with $t_i \in T_\mathcal{M}$.

As previously shown, the shape of our prediction function $run_\mathcal{S}$ is a PWLF with breakpoint in $\gamma$, \textit{i.e.},
\begin{equation}
run_\mathcal{S}(t) := \left\{\begin{array}{lr}
	\alpha                        & t \leq \gamma\\
	\beta\cdot(t-\gamma) + \alpha & t \geq \gamma\\
	\end{array}\right.
\label{eqn:p:lin}
\end{equation}

In order to train this function $run_\mathcal{S}(t)$, we search for those parameters $\alpha$, $\beta$, and $\gamma$ in \ref{eqn:p:lin} that minimise the percentage Root-Mean-Square Error (RMSE) on both the following sets.

Let $S_\mathcal{M}^{t<\gamma} := \lbrace(t_i,\tau_i) \in S_\mathcal{M} : t < \gamma\rbrace$ be the subset of $S_\mathcal{M}$ with $t$-values less than $\gamma$.
Similarly, let $S_\mathcal{M}^{t\geq\gamma} := \lbrace(t_i,\tau_i) \in S_\mathcal{M} : t \geq \gamma\rbrace$ be the subset of $S_\mathcal{M}$ with $t$-values greater or equal to $\gamma$.
We search for $\alpha$, $\beta$, and $\gamma$ that minimise the following sum.
\begin{equation}
Err_{S_\mathcal{M}} = \sum_{S \in \{\{S_\mathcal{M}^{t<\gamma}\}, \{S_\mathcal{M}^{t\geq\gamma}\}\}}\sqrt{\frac{1}{|S|} \sum_{(t_i,\tau_i) \in S} \left(\frac{\tau_i - run_\mathcal{S}(t_i)}{\tau_i}\right)^2}
\label{eqn:p:err}
\end{equation}

To this end, we use the CPLEX Optimiser\footnote{\url{https://www.ibm.com/products/ilog-cplex-optimization-studio}}. In particular, we iterate over different chosen values of $\gamma$, and let CPLEX decide the values of $\alpha$ and $\beta$ that minimise $Err_{S_\mathcal{M}}$ for the given $\gamma$. Finally, we choose those $\alpha$ and $\beta$ associated to the $\gamma$ with the minimum resulting $Err_{S_\mathcal{M}}$.

\clearpage
Table~\ref{fig:table:predictionfunc} shows the values $\alpha$, $\beta$, and $\gamma$ found using the training set $S_\mathcal{M}$, and the percentage RMSE $Err_{S_\mathcal{M}}$ associated to the function $run_\mathcal{S}$ for each Simulink model.

\begin{table}[H]
\begin{center}
\begin{tabular}[H]{|l|l|l|l|l|}
\hline
\hline
Model $\mathcal{M}$ &  $\alpha$ & $\beta$ & $\gamma$ & $Err_{S_\mathcal{M}}$\\
\hline
\hline
\texttt{sldemo\_fuelsys.mdl} & $0.39$ & $0.02$ & $0.4$ & $2.8\%$\\
\hline
\texttt{aero\_dap3dof.slx} & $0.39$ & $0.04$ & $0.1$ & $3.8\%$\\
\hline
\texttt{penddemo.slx} & $0.038$ & $0.0014$ & $10$ & $23.0\%$\\
\hline
\texttt{sldemo\_boiler.slx} & $0.1$ & $2\times10^{-5}$ & $50$ & $7.6\%$\\
\hline
\texttt{sldemo\_engine.slx} & $0.04$ & $0.01$ & $1.22$ & $2.0\%$\\
\hline
\texttt{sldemo\_househeat.slx} & $0.08$ & $0.0001$ & $10$ & $8.9\%$\\
\hline
\hline
\end{tabular}
\caption{Parameters of Prediction Function $run_\mathcal{S}$\label{fig:table:predictionfunc}}
\end{center}
\end{table}
\clearpage

\subsubsection{Other Commands}

Finally, we train our constants $load_\mathcal{S}$, $store_\mathcal{S}$, $free_\mathcal{S}$, and $inject_\mathcal{S}$ that
we need to define our final prediction function $\mathcal{P}^\mathcal{C}_\mathcal{S}$.
To this end, we compute the average execution time for each type of simulator command from the collected samples.
In particular, we set each prediction constant with the average execution time of the corresponding simulator command.

For example, let $S^{inject}_\mathcal{M} := (\tau_1, \tau_2, \ldots, \tau_K)$ be a training set that contains $K$ execution time samples collected by simulating \textit{inject} commands in $\mathcal{C}^*$. We define our constant $inject_\mathcal{S}$ as the average execution time of the gathered samples in $S^{inject}_\mathcal{M}$. Namely,
$$inject_\mathcal{S} := \frac{1}{|S^{inject}_\mathcal{M}|}\sum_{i \in [K]} \tau_i.$$
We do the same thing to train the other constants $load_\mathcal{S}$, $store_\mathcal{S}$, and $free_\mathcal{S}$ in $\mathcal{P}^\mathcal{C}_\mathcal{S}$.

\begin{figure}
	\begin{center}
		\includegraphics[width=\linewidth]{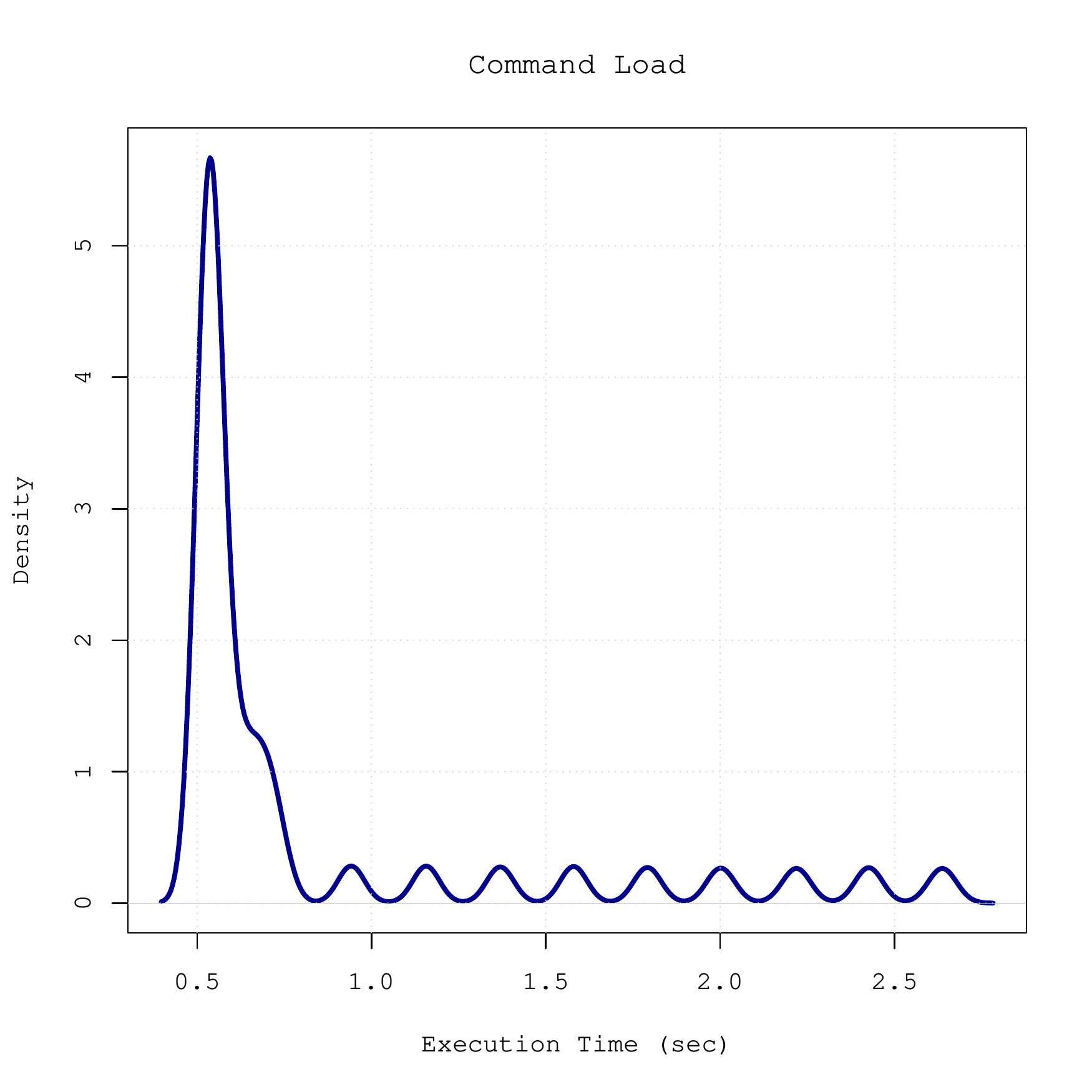}
		\caption{Execution Time Distribution of Command \textit{load}\label{fig:methodology:load:gaussian}}
	\end{center}
\end{figure}

\begin{figure}
	\begin{center}
		\includegraphics[width=\linewidth]{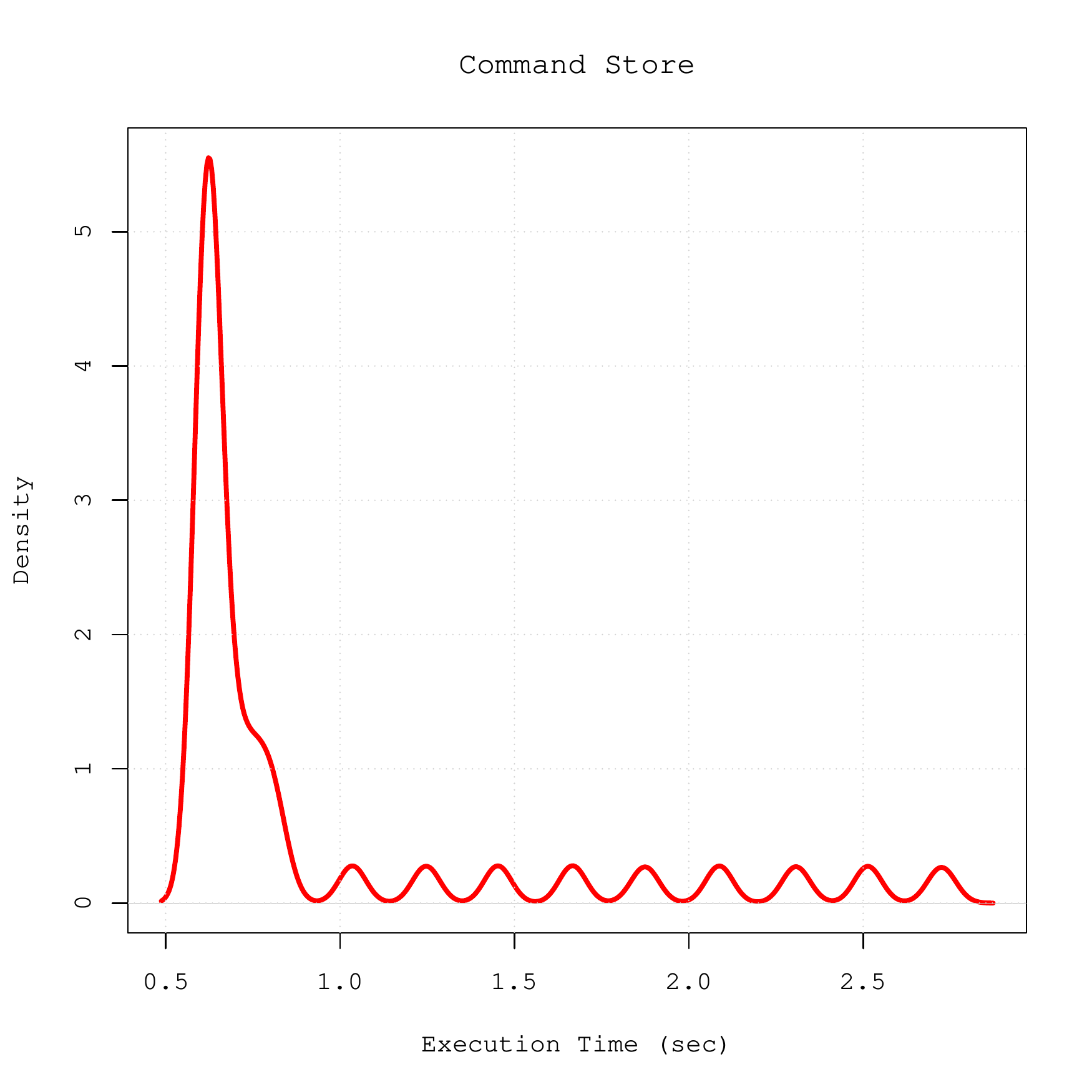}
		\caption{Execution Time Distribution of Command \textit{store}\label{fig:methodology:store:gaussian}}
	\end{center}
\end{figure}

\begin{figure}
	\begin{center}
		\includegraphics[width=\linewidth]{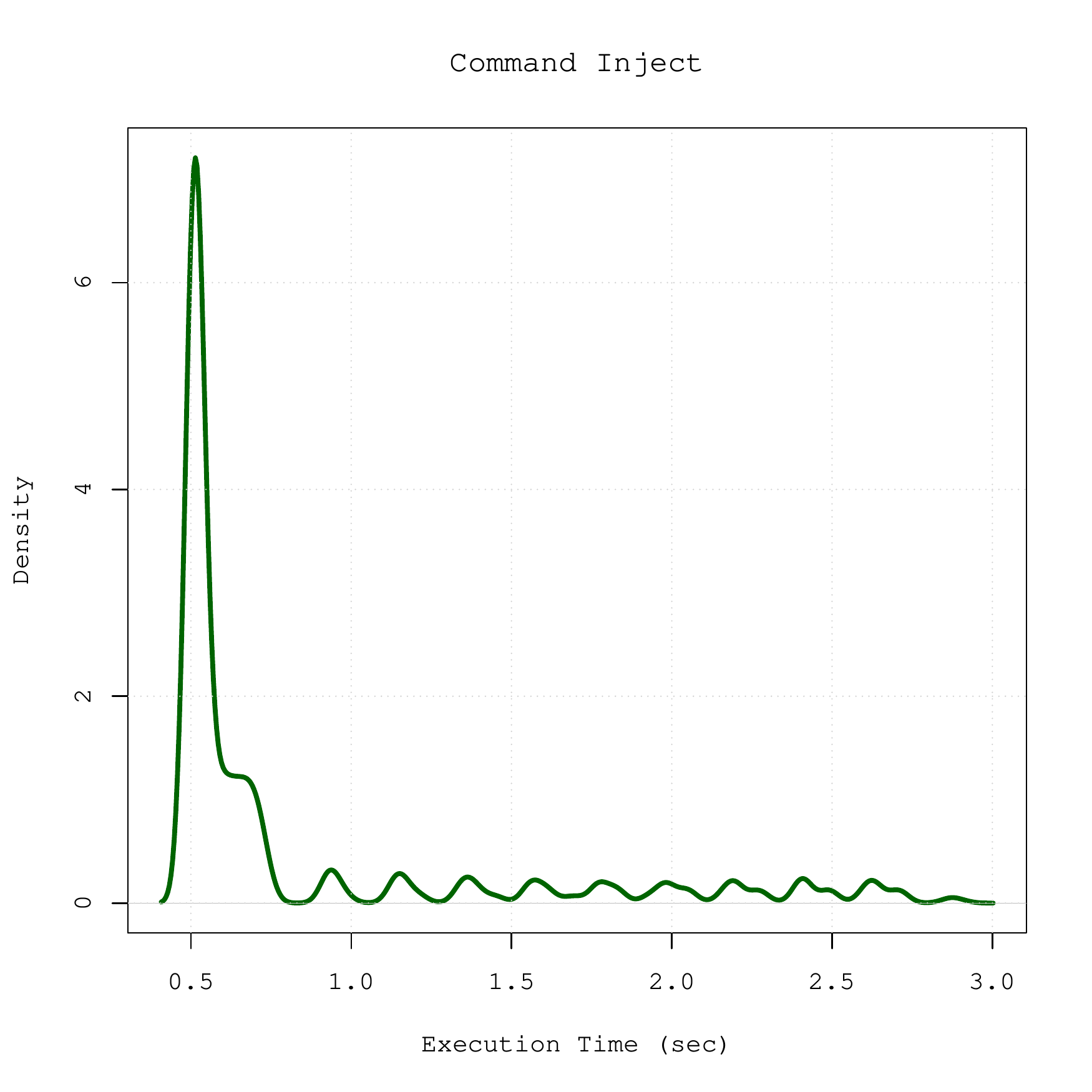}
		\caption{Execution Time Distribution of Command \textit{inject}\label{fig:methodology:inject:gaussian}}
	\end{center}
\end{figure}

\begin{figure}
	\begin{center}
		\includegraphics[width=\linewidth]{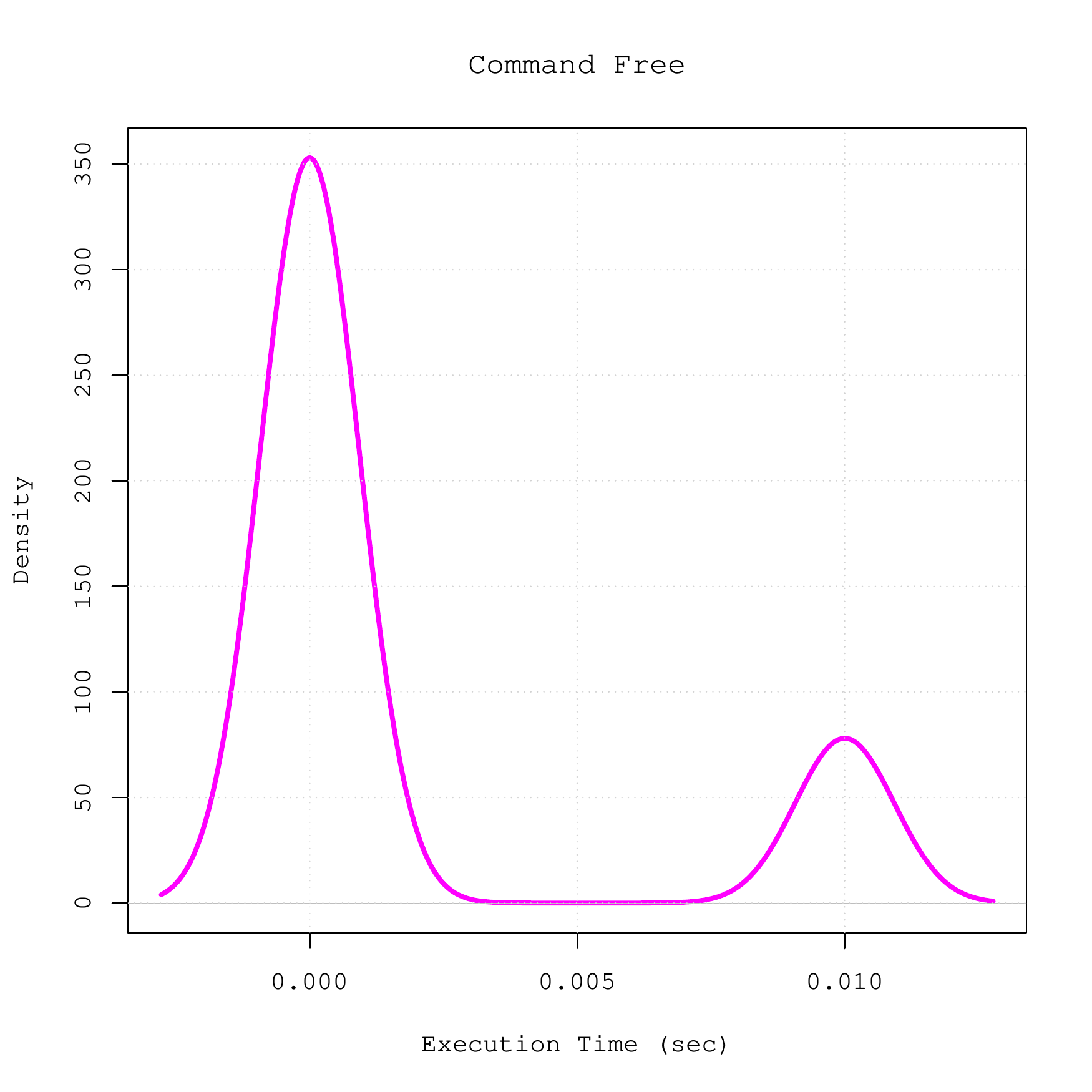}
		\caption{Execution Time Distribution of Command \textit{free}\label{fig:methodology:free:gaussian}}
	\end{center}
\end{figure}

The reason why we choose constants to estimate the execution time of these simulator commands is clearly shown in Figures~\ref{fig:methodology:inject:gaussian}, \ref{fig:methodology:store:gaussian}, \ref{fig:methodology:load:gaussian}, and \ref{fig:methodology:free:gaussian}. These figures show the distribution of the execution time samples of the corresponding simulator commands, that we collected by simulating $\mathcal{C}^*$ on the Fuel Control System model.

As an example, from Figure~\ref{fig:methodology:inject:gaussian} 
we see that the execution time to simulate commands $inject(d)$ in $\mathcal{C}^*$ is mostly between 0.5 and 0.6 seconds.

\begin{landscape}
\begin{figure}
	\begin{center}
		\includegraphics[width=1\linewidth]{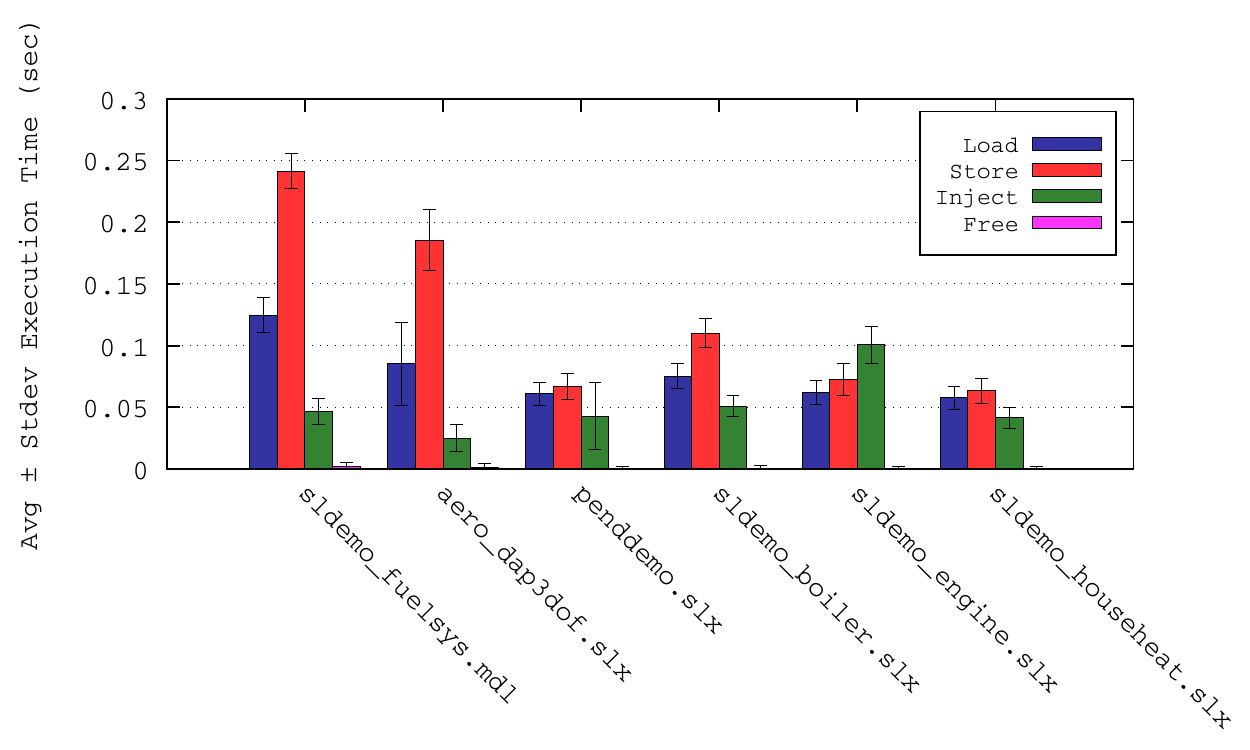}
		\caption{Execution Time for Commands
			\textit{load}, \textit{store}, \textit{inject}, and \textit{free}\label{fig:results:lsifavg}}
	\end{center}
\end{figure}
\end{landscape}

In conclusion, Figure~\ref{fig:results:lsifavg} shows the average and standard deviation of the execution time to run commands
\textit{load}, \textit{store}, \textit{inject}, and \textit{free}, on each Simulink model we used for our experiments.

\clearpage
\section{Experimental Results}

\subsection{Validation of the Method}

\subsubsection{Command Run}

Our goal is to validate the accuracy of the execution time estimate $run_\mathcal{S}(t)$ for any $t \in \mathbb{R}^+$.
For this reason, we specifically build a validation set for our prediction function $run_\mathcal{S}$ in the following way.
We define this set as $V^{run}_\mathcal{M} := \cup_{i=1}^K V_i$, where each $V_i$ is
$$V_i := \bigcup_{t \in T_\mathcal{M}} \lbrace (t', \tau) \rbrace\text{, with }t' \in [t\div10, t\cdot10].$$

Namely, we populate a meaningful number $K$ of validation sets $V_i$ by collecting execution time samples of $run(t)$ commands on the model $\mathcal{M}$.
In particular, we choose $t'$-values in each sample $(t', \tau)$ by picking a random value from the interval $[t\div10, t\cdot10]$, for every $t \in T_\mathcal{M}$.

Finally, we compute the prediction error of our function $run_\mathcal{S}$ as the average relative error between the measured execution time and the estimated one. Namely,
$$Err_{run_\mathcal{S}} := \frac{1}{|V^{run}_\mathcal{M}|} \sum_{(t,\tau) \in V^{run}_\mathcal{M}} \frac{\vert \tau-run_\mathcal{S}(x) \vert}{\tau}.$$

\begin{figure}
	\begin{center}
	\includegraphics[width=1\linewidth]{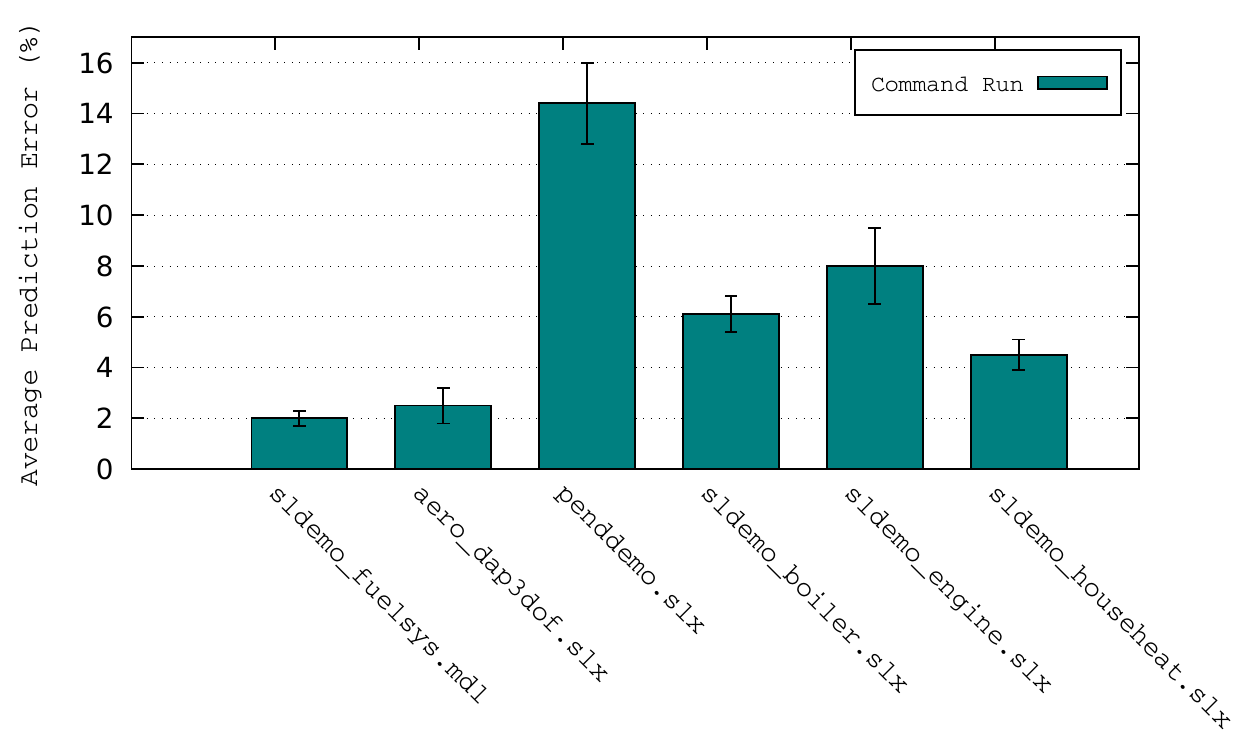}
	\caption{Average Prediction Error of $run_\mathcal{S}$\label{fig:results:runerr}}
	\end{center}
\end{figure}

Figure~\ref{fig:results:runerr} and Table~\ref{fig:table:runerr} show the average of the prediction errors found on each validation set $V_i$.
In particular, we obtain these results using $K=100$ validation sets for each Simulink model.

\begin{figure}
	\begin{center}
	\includegraphics[width=\linewidth]{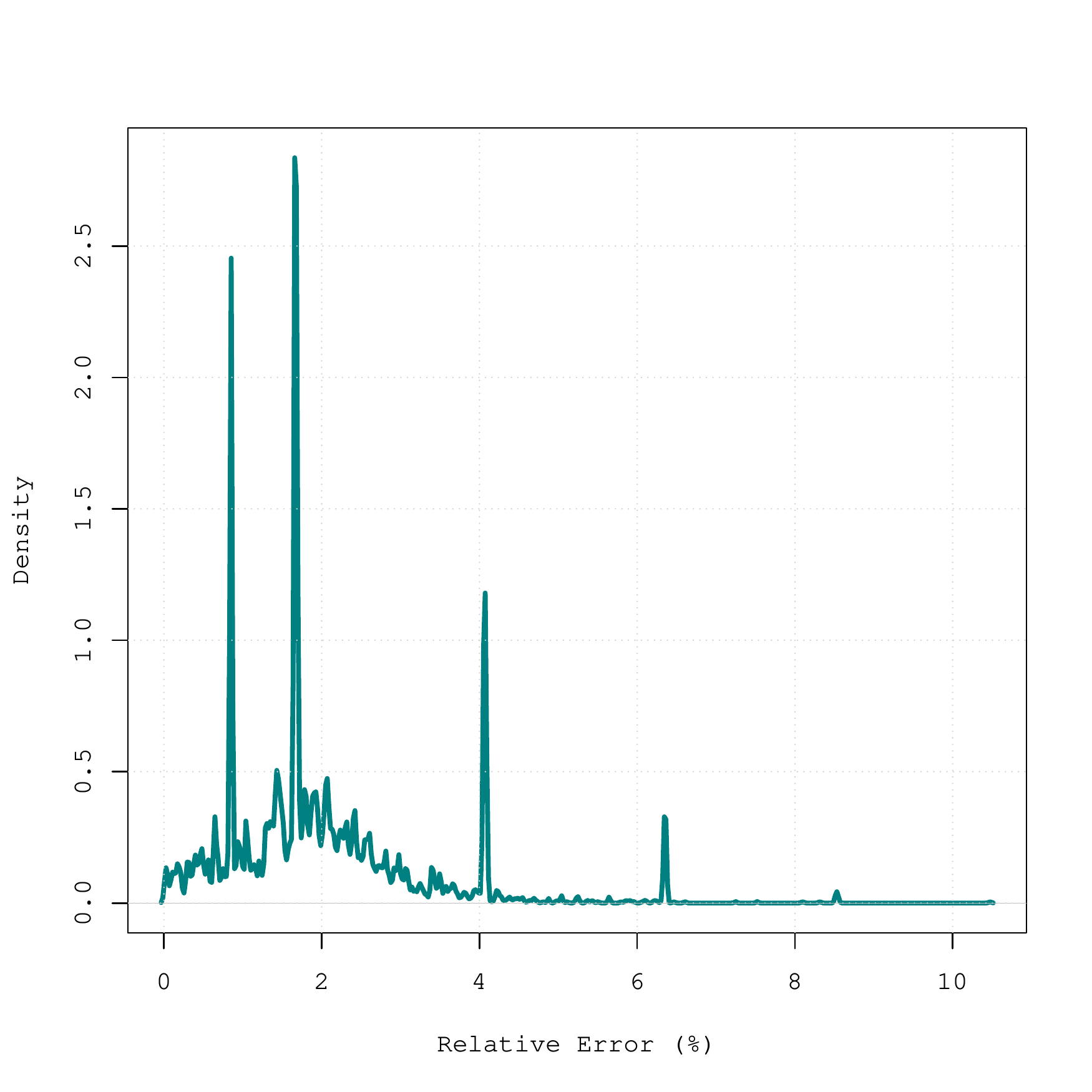}
	\caption{Relative Error Distribution of $run_\mathcal{S}$\label{fig:results:rundisterr}}
	\end{center}
\end{figure}

Figure~\ref{fig:results:rundisterr} shows the distribution of the relative prediction error of function $run_\mathcal{S}$ for the
Fuel Control System model, computed on the entire validation set $V^{run}_\mathcal{M}$.

\renewcommand{\arraystretch}{1.2}
\begin{table}
\begin{center}
\resizebox{\textwidth}{!}{
\begin{tabular}[H]{|l|l|l|l|l|}
\hline
\hline
\textsc{Model }$\mathcal{M}$ &  Minimum & Maximum & Average & Standard Deviation\\
\hline
\hline
\texttt{sldemo\_fuelsys.mdl} & 1.1 & 2.5 & 2.0 & 0.3\\
\hline
\texttt{aero\_dap3dof.slx} & 1.9 & 5.5 & 2.5 & 0.7\\
\hline
\texttt{penddemo.slx} &10.9 & 18.3 & 14.4 & 1.6\\
\hline
\texttt{sldemo\_boiler.slx} & 3.9 & 7.6 & 6.1 & 0.7\\
\hline
\texttt{sldemo\_engine.slx} & 5.7 & 10.6 & 8.0 & 1.5\\
\hline
\texttt{sldemo\_househeat.slx} & 3.4 & 5.7 & 4.5 & 0.6\\
\hline
\hline
\end{tabular}}
\caption{Average Prediction Error of $run_\mathcal{S}$\label{fig:table:runerr}}
\end{center}
\end{table}

\clearpage
\subsubsection{Simulation Campaigns}
\vspace*{-10pt}
\renewcommand{\arraystretch}{1.3}
\newcolumntype{Y}{>{\centering\arraybackslash}X}
\begin{table}[H]
\begin{center}
\begin{tabularx}{1.1\textwidth}{|*{6}{Y|}}
\hline
\hline
\multirow{2}{*}{\textsc{Slices}} &
	\multicolumn{3}{c|}{\textsc{Average per Slice (hh:mm:ss)}} &
	\multicolumn{2}{c|}{\textsc{Relative Error (\%)}}\\
\cline{2-6} &
	\footnotesize Measured &
	\footnotesize Estim\textsubscript{$\alpha\beta\gamma$} &
	\footnotesize Estim\textsubscript{na{\"i}ve} &
	\footnotesize Err\textsubscript{$\alpha\beta\gamma$} &
	\footnotesize Err\textsubscript{na{\"i}ve}\\
\hline
\hline
	128 & 96:08:22 & 99:23:49 & 99:32:35 & 7.09 & 7.07\\
\hline
	256 & 51:36:56 & 51:38:26 & 51:49:34 & 1.29 & 1.20\\
\hline
	512 & 26:04:26 & 26:08:05 & 26:13:25 & 2.59 & 2.43\\
\hline
\hline
\end{tabularx}
\caption{Estimation Error of Simulation Campaings\label{fig:table:simcamp-err}}
\end{center}
\end{table}

Table~\ref{fig:table:simcamp-err} describes the estimation results from our experiments.
It illustrates both the measured and estimated execution time to simulate slices of simulation scenarios.
These slices are obtained from a dataset of scenarios that we generated automatically from a
formal model of disturbances to verify the
\textit{Apollo}\footnote{\url{https://it.mathworks.com/help/simulink/examples/developing-the-apollo-lunar-module-digital-autopilot.html}} Simulink model.

To perform these experiments, we first split the initial dataset of scenarios into 128, 256, and 512 slices.
Then, from these slices we compute the corresponding optimised simulation campaigns.
Finally, we execute each simulation campaign in parallel, and collect the elapsed time of the simulation.

Results from Table~\ref{fig:table:simcamp-err} show that the measured
elapsed time to run a simulation campaign from the 128-slice group is around 96~hours on average.
The corresponding elapsed time estimated with both our presented methods~(\textit{i.e.},
Estim\textsubscript{$\alpha\beta\gamma$}) and the na{\"i}ve one (\textit{i.e.},
Estim\textsubscript{na{\"i}ve}) is around 99~hours on average, with an error  of approximately 7\%.

It is worth noting that the na{\"i}ve estimation was obtained from
execution time samples collected by running an entire simulation campaign from the 128-slice group.
The execution of this campaign as a whole required around 96~hours of simulation.

In contrast, our estimation was obtained from just a few execution time samples collected by running a small simulation campaign we computed ad-hoc. In particular, this simulation campaign is made of about 100 simulator instructions. As can be seen, the execution of this campaign  was significantly shorter and is measured in terms of minutes, rather than hours. In fact the total amount of time required was a mere 10~minutes of simulation.

\clearpage
\section{Related Work}

To the best of our knowledge, no previous work in the literature directly addresses the challenge of estimating the execution time of simulation campaigns~\cite{McHaney1991}. In fact, when cluster usage is limited to a maximum walltime $w$ for computation tasks, the typical workaround used for computations requiring more than $w$ seconds is called \textit{checkpointing}~\cite{Walters2009}. Namely, the computation task must periodically save its state (checkpoint). In this way, if the task is killed--either because the walltime has expired, or because the cluster has faulted~\cite{Walters2009}--it may be restarted from the last checkpoint rather than from the beginning.

The use of checkpointing is not appropriate in our setting for the following reasons:
(1)~The application code must be instrumented in order to perform checkpointing. Although this is leveraged by existing software libraries~\cite{Bautista2011}, it requires both access to and knowledge of the source code, which may not be possible.
(2)~Checkpointing does not provide trade-off between computational resources and splitting of simulation campaigns into smaller or bigger slices, which is our focus here.

With regard to hardware/software Worst Case Execution Time (WCET), some research has been done on studying algorithms and methods to estimate it. As an example, in~\cite{Tan2009} a systematic method is presented that makes model information available for timing analysis and presents promising results with Simulink/Stateflow models.

In~\cite{Zolda2016} an approach based on integer linear programming is presented for calculating a WCET estimate from a given database of timed execution traces.

The main differences between estimation of WCET and estimation of the execution time of simulation campaigns are the following:
(1)~WCET aims to determine the worst scenario of execution of a system, mainly to check that hard real-time requirements in hardware/software interactions are met. On the other hand, our focus here is on the {\em average} execution time of a simulation campaign.
(2)~WCET is typically tailored to some specific hardware architecture. On the contrary, our method targets any computer architecture, as it also contains a hardware-dependent training phase.

\clearpage
\section{Conclusions and Future Work}

\subsection{Conclusions}

In this chapter, we presented an effective method to perform an accurate execution time estimate of simulation campaigns.
In particular, we described our machine-independent approach to selecting a small number of simulators commands to 
collect execution time samples from. Furthermore, we showed how to train a prediction function from the collected samples.

Results show that our method can effectively predict the time needed to execute a simulation campaign with
an error below~10\%.

%% file: chap05.tex
\chapter{Conclusions and Future Work}\label{chap:conclusions}

\section{Conclusions}

In this thesis, we have presented an optimiser and an execution time estimator employed to overcome the limitations associated with 
Model Based System-Level Formal Verification (SLFV) tools for complex Cyber-Physical Systems (CPSs). Namely the significant amount of time required for simulations as well as the unpredictability of  its length, which prevent a more widespread adoption of SLFV  tools over the entire development life cycle.

In Chapter~\ref{chap:optimiser}, we presented an efficient data-intensive optimiser tool to speed up the SLFV process by
obtaining optimised simulation campaigns from the existing datasets of scenarios to be verified.
Optimised simulation campaigns accelerate SBV by exploiting the capability of modern simulators to store and re-use intermediate simulation states in order to eliminate the need to explore common paths of scenarios multiple times.

In Chapter~\ref{chap:estimator}, we presented an effective partially machine-independent execution time estimator tool to predict the 
required time to complete SLFV activities.
An accurate estimate of execution time leads to better planning of deadlines, and wiser budget allocation for the required
computing resources.

\clearpage
\section{Future Work}

Experimental results in Section~\ref{sec:results-optimiser} indicate that our optimisation method could also be scaled horizontally.
In fact, Figure~\ref{fig:results:wallclock-steps} shows how the sorting step is significantly more time-consuming than all the others. In fact, it takes as much as 12 hours to sort the 4-TB input dataset, 6 times longer than the other steps put together.

For this reason, an interesting further investigation could be to delegate this step to a cluster-computing framework such as
Apache Spark\footnote{\url{https://spark.apache.org/}} in a dedicated cluster. Such an approach has indeed led to very promising results in recent publications. For example, a team from the Apache Spark community showed how they managed to sort 100~TB of data 
in as little as 23 minutes \cite{Armbrust2015}.

Another investigation into the applicability of the optimiser tool would be the optimisation of existing 
datasets used in CPS companies. This would require the implementation of the following pre-processing step
in order to convert existing scenarios into the input format that we use for our optimiser.
Faulty events in the existing dataset would be encoded with disturbance IDs (\textit{i.e.}, disturbances). Next, the fixed length $H$ of scenarios that we use in our input format could be obtained from the existing dataset by searching for the largest sequence of disturbances in the same existing dataset. Once $H$ is obtained, all the other scenarios with length less than $H$ in the existing dataset could be expanded by appending to them the appropriate number of zeros (\textit{i.e.}, non-disturbances) in order to make all the resulting scenarios the same length $H$.

Since there is currently a great interest about continuous integration and deployment of new software components in CPSs that are already in use, \textit{e.g.}, new subsystems in vehicles that are already on the road, it would be interesting to investigate on the possibility to use the solutions presented in this thesis together with incremental verification techniques~\cite{IncrementalVer2013} in order to select only those simulation scenarios that are actually needed to be re-verified, according to the performed software changes.

In conclusion, while in this thesis we focused on Simulink to devise our estimation method, it would be interesting to investigate the applicability of our method with other industrially viable simulators such as
Ngspice\footnote{\url{http://ngspice.sourceforge.net/}} and 
JModelica\footnote{\url{http://www.jmodelica.org/}}.
The use of these simulators would require the implementation of two driver tools to translate each simulator command from the simulation campaign 
into the corresponding instructions to be executed by Ngspice and JModelica. A Simulink implementation of such a driver tool can be found at our Bitbucket repository\footnote{\url{https://bitbucket.org/mclab/}}.

%% file: appendix.tex
\addappheadtotoc
\chapter{Optimiser Tool: Implementation and Usage}

In this section we show how we implemented our optimiser tool.
In particular, we first illustrate both the input and output format,
and then we describe each step of the algorithm.

The implemented code of the presented optimiser tool
is available at our Bitbucket repository\footnote{\url{https://bitbucket.org/mclab/dt-optimiser/}}.

\section{Definitions}

\subsection*{Lex-Ordered Dataset of Disturbance Traces}

A lex-ordered dataset of Disturbance Traces (DTs) is an ordered sequence $\mathcal{D}$
that contains $|\mathcal{D}| = N$ distinct DTs with length $H$, \textit{i.e.},
$$\mathcal{D} := \bigcup_{i \in [N]} \{ (d^i_1,d^i_2,\ldots,d^i_H) \}.$$
In particular, each disturbance $d^i_j \in \mathbb{N}^+ \cup \{0\}$ is a natural number that encodes a fault to inject 
on the model being simulated. Furthermore, zeros indicate non-disturbances.

\subsection*{Simulation Scenario}

Given the $i$-th DT $\delta_i = (d^i_1,d^i_2,\ldots,d^i_H)$ in a lex-ordered dataset $\mathcal{D}$,
a \textit{simulation scenario} (or \textit{scenario}) is a sequence of $H$ consecutive simulation intervals
where we inject the model being simulated with each disturbance $d^i_j \in \delta_i$ at the beginning of
the corresponding $j$-th interval. Note that if $d^i_j = 0$, then no disturbance is injected.
Furthermore, each simulation interval has a fixed length $\tau$ of simulation seconds.
In conclusion, we give the name $H$ to the \textit{simulation horizon} (or \textit{horizon}).

\subsection*{Simulation Campaign}

Given a lex-ordered dataset $\mathcal{D}$ of DTs in input,
a simulation campaign is a sequence of \textit{simulator commands} that are aimed at
simulating scenarios in $\mathcal{D}$.
Simulator commands are described in Table~\ref{fig:table:appendix:commands} below.

\subsection*{Simulator Commands}

Table~\ref{fig:table:appendix:commands} shows the syntax and behaviour of the five basic simulator commands.

\renewcommand{\arraystretch}{1.2}
\begin{table}[H]
\begin{center}
\resizebox{1.05\textwidth}{!}{
\begin{tabular}[H]{r|l}
\textsc{Syntax} & \textsc{Behaviour}\\
\hline
$\text{{\tt I}\textless}\text{{\tt int}\textgreater}$ & {\em injects the disturbance $\text{\textless}\text{{\tt int}\textgreater}$ on the model being simulated}\\
\hline
$\text{{\tt R}\textless}\text{{\tt int}\textgreater}$ & {\em advances the simulator state by $\text{\textless}\text{{\tt int}\textgreater}$ simulation intervals}\\
\hline
$\text{{\tt S}\textless}\text{{\tt int}\textgreater}$ & {\em saves the current simulator state in a file named $\text{\textless}\text{{\tt int}\textgreater}$}\\
\hline
$\text{{\tt L}\textless}\text{{\tt int}\textgreater}$ & {\em loads the simulator state from the file $\text{\textless}\text{{\tt int}\textgreater}$}\\
\hline
$\text{{\tt F}\textless}\text{{\tt int}\textgreater}$ & {\em removes the file $\text{\textless}\text{{\tt int}\textgreater}$}\\
\end{tabular}}
\caption{Syntax of Simulator Commands\label{fig:table:appendix:commands}}
\end{center}
\end{table}
 
\section{Input Format}\label{input-format}

The input file (DT file) is a binary file that contains
DTs where disturbances are encoded by unsigned 64-bit integers.

In particular, a DT file contains a multiset of DTs that have horizon $H$, thus DTs are not lex-ordered so there can be duplicate DTs.
Furthermore, there is no header in the DT file that indicates the horizon H of DTs. Also, there is no separator between DTs.

Hence, a DT file is a simple sequence of $N\cdot H$ disturbances, where $N$ is the number of DTs in the DT file, and $H$ is their horizon.

\textbf{Example A.2} (Input File). Below is an example of an input DT file.
Specifically, it contains a multiset of DTs with horizon $H = 5$.
Note that DTs are not lex-ordered and there are duplicate DTs.

\begin{tcolorbox}[width=\textwidth,colback=white,arc=0pt,bottom=5pt,boxrule=0.3pt]
{\small
\begin{Verbatim}[commandchars=\\\{\}]
\colorbox{white}{0} \colorbox{white}{0} \colorbox{white}{0} \colorbox{white}{0} \colorbox{white}{0}
\colorbox{blue!20}{1} \colorbox{white}{0} \colorbox{white}{0} \colorbox{blue!20}{1} \colorbox{white}{0}
\colorbox{white}{0} \colorbox{white}{0} \colorbox{blue!20}{1} \colorbox{blue!20}{2} \colorbox{white}{0}
\colorbox{white}{0} \colorbox{white}{0} \colorbox{blue!20}{1} \colorbox{white}{0} \colorbox{white}{0}
\colorbox{white}{0} \colorbox{blue!20}{1} \colorbox{white}{0} \colorbox{white}{0} \colorbox{blue!20}{1}
\colorbox{blue!20}{1} \colorbox{white}{0} \colorbox{white}{0} \colorbox{white}{0} \colorbox{white}{0}
\colorbox{blue!20}{1} \colorbox{white}{0} \colorbox{white}{0} \colorbox{white}{0} \colorbox{white}{0}
\end{Verbatim}
}
\end{tcolorbox}

For the sake of clarity, the content of the input DT file above is shown 
in a textual format. Namely, disturbances are shown in their textual representation
and are separated by space characters. Furthermore, DTs are separated by new lines.

\section{Output Format}\label{output-format}

The output file (optimised simulation campaign) is a text file with simulator commands
that are aimed at simulating the corresponding DTs in the input lex-ordered dataset.

Specifically, each line in the optimised simulation campaign (say the $i$-th line) contains simulator commands
that simulate the corresponding DT in input (the $i$-th DT in the dataset).
Each line has the following syntax.
{\small
$$\text{{\tt F}\textless}\text{{\tt int}\textgreater}\{0,H-1\}\text{ }\text{{\tt L}\textless}\text{{\tt int}\textgreater}\{1\}\text{ }(\text{{\tt I}\textless}\text{{\tt int}\textgreater}\{0,1\}\text{ }\text{{\tt R}\textless}\text{{\tt int}\textgreater}\{1\}\text{ }\text{{\tt S}\textless}\text{{\tt int}\textgreater}\{0,1\})\{1,H\}$$}\\[-10pt]
In particular, there can be a number from 0 to $H-1$ of commands \textit{free}
(\textit{i.e.}, $\text{{\tt F}\textless}\text{{\tt int}\textgreater}$), depending on
the current number of previously stored simulation states that are no longer needed
by future DTs.

Then, there is the command \textit{load} (\textit{i.e.}, $\text{{\tt L}\textless}\text{{\tt int}\textgreater}$).
Note that this command is mandatory. In fact, each input DT in the resulting simulation campaign starts from
its corresponding initial state. Note that $\text{\textless}\text{{\tt int}\textgreater}$ indicates an integer.

Lastly, there is a number from 1 to $H$ of \textit{inject-run-store} commands
(\textit{i.e.}, $\text{{\tt I}\textless}\text{{\tt int}\textgreater}\{0,1\}\text{ }\text{{\tt R}\textless}\text{{\tt int}\textgreater}\{1\}\text{ }\text{{\tt S}\textless}\text{{\tt int}\textgreater}\{0,1\}$), depending on both the number of disturbances to inject
and the number of states to store. Specifically, commands \textit{inject} and \textit{store}
(\textit{i.e.}, $\text{{\tt I}\textless}\text{{\tt int}\textgreater}$, and $\text{{\tt S}\textless}\text{{\tt int}\textgreater}$)
are optional since there can be neither disturbances to inject nor states to store at certain simulation intervals.

\textbf{Example A.3} (Output File). Below is an example of an optimised simulation campaign
that corresponds to the lex-ordered dataset of input DTs shown in the input file example.

\begin{tcolorbox}[width=\textwidth,colback=white,arc=0pt,bottom=5pt,boxrule=0.3pt]
{\small
\begin{Verbatim}[commandchars=\\\{\}]
L0 R1 S1 R1 S2 R3
L2 I\colorbox{blue!20}{1} R1 S8 R2
L8 I\colorbox{blue!20}{2} R2
F2 F8 L1 I\colorbox{blue!20}{1} R3 I\colorbox{blue!20}{1} R1
F1 L0 I\colorbox{blue!20}{1} R3 S23 R2
L23 I\colorbox{blue!20}{1} R2
F23
\end{Verbatim}
}
\end{tcolorbox}

Note that the command {\tt L}0 loads the initial simulator state, which is
assumed to exist prior to the simulation, thus there is no previous {\tt S}0 command
in the simulation campaign.

\clearpage
\section{Algorithm Steps}

\subsection*{Step 1. Initial Sorting}
\begin{description}
	\setlength\itemsep{0em}
	\item[INPUT:] A file that contains a multiset of DTs with length $H$.
	\item[OUTPUT:] A file that contains the corresponding lex-ordered dataset $\mathcal{D}$~of~DTs.
\end{description}

The output dataset $\mathcal{D}$ is obtained from uniquely sorting the input DTs.

\textbf{Example A.4.1} (Step 1). The example below shows the sequence of DTs in input
(\textit{i.e.}, from the Example~\ref{input-format}) and the resulting lex-ordered
dataset $\mathcal{D}$ of DTs in output.

\begin{tcolorbox}[width=\textwidth,colback=white,arc=0pt,bottom=5pt,boxrule=0.3pt]
\begin{Verbatim}[commandchars=\\\{\}]
\colorbox{gray!20}{Input}             \colorbox{gray!20}{Output}

\colorbox{white}{0} \colorbox{white}{0} \colorbox{white}{0} \colorbox{white}{0} \colorbox{white}{0}     \colorbox{white}{0} \colorbox{white}{0} \colorbox{white}{0} \colorbox{white}{0} \colorbox{white}{0}
\colorbox{blue!20}{1} \colorbox{white}{0} \colorbox{white}{0} \colorbox{blue!20}{1} \colorbox{white}{0}     \colorbox{white}{0} \colorbox{white}{0} \colorbox{blue!20}{1} \colorbox{white}{0} \colorbox{white}{0}
\colorbox{white}{0} \colorbox{white}{0} \colorbox{blue!20}{1} \colorbox{blue!20}{2} \colorbox{white}{0}     \colorbox{white}{0} \colorbox{white}{0} \colorbox{blue!20}{1} \colorbox{blue!20}{2} \colorbox{white}{0}
\colorbox{white}{0} \colorbox{white}{0} \colorbox{blue!20}{1} \colorbox{white}{0} \colorbox{white}{0}     \colorbox{white}{0} \colorbox{blue!20}{1} \colorbox{white}{0} \colorbox{white}{0} \colorbox{blue!20}{1}
\colorbox{white}{0} \colorbox{blue!20}{1} \colorbox{white}{0} \colorbox{white}{0} \colorbox{blue!20}{1}     \colorbox{blue!20}{1} \colorbox{white}{0} \colorbox{white}{0} \colorbox{white}{0} \colorbox{white}{0}
\colorbox{blue!20}{1} \colorbox{white}{0} \colorbox{white}{0} \colorbox{white}{0} \colorbox{white}{0}     \colorbox{blue!20}{1} \colorbox{white}{0} \colorbox{white}{0} \colorbox{blue!20}{1} \colorbox{white}{0}
\colorbox{blue!20}{1} \colorbox{white}{0} \colorbox{white}{0} \colorbox{white}{0} \colorbox{white}{0}
\end{Verbatim}
\end{tcolorbox}

\subsection*{Step 2. Load Labelling}
\begin{description}
	\setlength\itemsep{0em}
	\item[INPUT:] A file that contains the lex-ordered dataset of DTs $\mathcal{D}$ computed at Step 1.
	\item[OUTPUT:] A file with the sequence of \textit{load labels} $\mathcal{L} := (\mathcal{L}_1, \mathcal{L}_2, \ldots, \mathcal{L}_N)$.
\end{description}

In order to describe the labelling strategy in more detail, let first define $p(i)$ as the size of the 
Longest Common Prefix (LCP) between the $i$-th and the $(i-1)$-th DTs in $\mathcal{D}$, \textit{i.e.}, 
$$p(i) := | \{ p \in [H] : \bigwedge\limits_{j = 1}^p d^i_j = d^{i-1}_j \} |.$$

In other words, $p(i)$ represents the index of the rightmost disturbance of the LCP. Note that $p(i) = 0$ if no common prefix exists.

\clearpage
\subsubsection*{LABELING STRATEGY}
First of all, disturbances in the 1-st DT are assigned with a label $\ell^1_j = j$
for each $j \in \{1, 2, \ldots, H\}$. For all the other DTs, each disturbance $d^i_j$ is assigned 
a label $\ell^i_j$ as follows.
$$\ell^i_j := \left\{\begin{array}{lr}
        \ell^{i-1}_j        & \text{if }j \leq p(i)\\
        (i-1)\cdot H + j    & \text{if }j > p(i)\\
        \end{array}\right.$$

Note that to compute each label $\ell^i_1, \ell^i_1, \ldots$, and $\ell^i_H$ there is no need to
load $\mathcal{D}$ entirely into the main memory. In fact, all we need is to compare the $i$-th DT in $\mathcal{D}$ with the previous one.

\textbf{Final Sequence.} The final sequence of \textit{load labels} $\mathcal{L} := (\mathcal{L}_1, \mathcal{L}_2, \ldots, \mathcal{L}_N)$
is computed in such a way that, for each $i \in \{1,2,\ldots,N\}$,
$$\mathcal{L}_i := \left\{\begin{array}{lr}
        \ell^i_{p(i)} & \text{if }p(i) > 0\\
        0             & \text{if }p(i) = 0\\
        \end{array}\right.$$

\textbf{Example A.4.2} (Step 2). The example below shows the input dataset $\mathcal{D}$ of
lex-ordered DTs (from Step 1), and the output sequence of \textit{load labels}.

\begin{tcolorbox}[width=\textwidth,colback=white,arc=0pt,bottom=5pt,boxrule=0.3pt]
\begin{Verbatim}[commandchars=\\\{\}]
\colorbox{gray!20}{Input} (Sorted DTs)         \colorbox{gray!20}{Output} (Load Labels)

\colorbox{white}{0} \colorbox{white}{0} \colorbox{white}{0} \colorbox{white}{0} \colorbox{white}{0}              \colorbox{orange!60}{0}
\colorbox{white}{0} \colorbox{white}{0} \colorbox{blue!20}{1} \colorbox{white}{0} \colorbox{white}{0}              \colorbox{orange!60}{2}
\colorbox{white}{0} \colorbox{white}{0} \colorbox{blue!20}{1} \colorbox{blue!20}{2} \colorbox{white}{0}              \colorbox{orange!60}{8}
\colorbox{white}{0} \colorbox{blue!20}{1} \colorbox{white}{0} \colorbox{white}{0} \colorbox{blue!20}{1}              \colorbox{orange!60}{1}
\colorbox{blue!20}{1} \colorbox{white}{0} \colorbox{white}{0} \colorbox{white}{0} \colorbox{white}{0}              \colorbox{orange!60}{0}
\colorbox{blue!20}{1} \colorbox{white}{0} \colorbox{white}{0} \colorbox{blue!20}{1} \colorbox{white}{0}              \colorbox{orange!60}{23}
\end{Verbatim}
\end{tcolorbox}

Note that the output sequence of \textit{load labels} is stored in a file in the same format as the input DTs.
Namely, each label is a 64-bit unsigned integer. Furthermore, there is no separator between labels.

For the sake of clarity, the output sequence above is shown in a textual format.

\clearpage
\subsection*{Step 3. Store Labelling}

\begin{description}
	\setlength\itemsep{0em}
	\item[INPUT:] A file that contains the sequence of \textit{load labels} $\mathcal{L}$ computed~at~Step~2.
	\item[OUTPUT:] A file that contains the resulting lex-ordered sequence of \textit{store labels} $\mathcal{S}$.
\end{description}

The output sequence is obtained from uniquely sorting the input sequence.
The label \colorbox{brown}{\textcolor{white}{0}} in the resulting sequence is then removed.
In fact, since $0$ represents the initial simulator state, there is no need to store it.\\

\textbf{Example A.4.3} (Step 3). The example below shows the input sequence of \textit{load labels},
and the output sequence of \textit{store labels}.
\begin{tcolorbox}[width=\textwidth,colback=white,arc=0pt,bottom=5pt,boxrule=0.3pt]
\begin{Verbatim}[commandchars=\\\{\}]
\colorbox{gray!20}{Input} (Load Labels)         \colorbox{gray!20}{Output} (Store Labels)

\colorbox{orange!60}{0}                           \colorbox{brown}{\textcolor{white}{1}}
\colorbox{orange!60}{2}                           \colorbox{brown}{\textcolor{white}{2}}
\colorbox{orange!60}{8}                           \colorbox{brown}{\textcolor{white}{8}}
\colorbox{orange!60}{1}                           \colorbox{brown}{\textcolor{white}{23}}
\colorbox{orange!60}{0}
\colorbox{orange!60}{23}
\end{Verbatim}
\end{tcolorbox}

\subsection*{Step 4. Final Optimisation}
\begin{description}
	\setlength\itemsep{0em}
	\item[INPUT 1:] The file that contains $\mathcal{D}$ computed at Step 1.
	\item[INPUT 2:] The file that contains $\mathcal{L}$ computed at Step 2.
	\item[INPUT 3:] The file that contains $\mathcal{S}$ computed at Step 3.
	\item[OUTPUT:] The final optimised simulation campaign.
\end{description}

Each line in the computed simulation campaigns (say the $i$-th line) consists of simulator commands that simulate
the corresponding DT (the $i$-th one) from the input lex-ordered dataset of DTs.

\clearpage
\subsubsection*{DETAILED DESCRIPTION OF STEP 4}

As previously stated, each line in the optimised simulation campaign has the following syntax.
{\small
$$\text{{\tt F}\textless}\text{{\tt int}\textgreater}\{0,H-1\}\text{ }\text{{\tt L}\textless}\text{{\tt int}\textgreater}\{1\}\text{ }(\text{{\tt I}\textless}\text{{\tt int}\textgreater}\{0,1\}\text{ }\text{{\tt R}\textless}\text{{\tt int}\textgreater}\{1\}\text{ }\text{{\tt S}\textless}\text{{\tt int}\textgreater}\{0,1\})\{1,H\}$$}\\[-15pt]
Hence, for each $i \in \{1,2,\ldots,N\}$, we output a textual line of simulator commands in the following way.

First of all, let $\mathcal{F} := (\mathcal{F}_0,\mathcal{F}_2,\ldots,\mathcal{F}_{H-1})$ be defined as the {\em array for commands free}.
In particular, $\mathcal{F}$ is an array with $H$ elements initialised to zero. As the name suggests,
this array helps us spot those previously stored labels that are no longer required.

\begin{description}
	\item[Part $\text{{\tt F}\textless}\text{{\tt int}\textgreater}\{0,H-1\}$]\hfill\\[3pt]
	For each $j \in \{p(i),p(i)+1,\ldots,H-1\}$,
	if $\mathcal{F}_j > 0$, then $\text{{\tt print F}}\mathcal{F}_j$, and then set the $j$-th position of the array to zero,
	\textit{i.e.}, $\mathcal{F}_j \gets 0.$

	\item[Part $\text{{\tt L}\textless}\text{{\tt int}\textgreater}\{1\}$]\hfill\\[3pt]
	First, $\text{{\tt print L}}\mathcal{L}_i$,
	and put the label  $\mathcal{L}_i$ into the $p(i)$-th position of the array $\mathcal{F}$,
	\textit{i.e.}, $\mathcal{F}_{p(i)} \gets \mathcal{L}_i$.

	\item[Part $(\text{{\tt I}\textless}\text{{\tt int}\textgreater}\{0,1\}\text{ }\text{{\tt R}\textless}\text{{\tt int}\textgreater}\{1\}\text{ }\text{{\tt S}\textless}\text{{\tt int}\textgreater}\{0,1\})\{1,H\}$]\hfill\\[3pt]
	Let $\mathcal{I}:=\mathcal{I}_1,\mathcal{I}_2,\ldots,\mathcal{I}_k$
	be a sequence of $k$ indexes such that \mbox{$\mathcal{I}_1 = p(i)+1$}, $\mathcal{I}_k = H+1$, and for all the other
	$\mathcal{I}_j$, with $\mathcal{I}_1 < \mathcal{I}_j < \mathcal{I}_k$, 
	there is either a disturbance
	$d^i_{\mathcal{I}_j}$ to inject, or there is a label to store, in other words the $(\mathcal{I}_j-1)$-th label in
	the current DT, \textit{i.e.}, $(d^i_{\mathcal{I}_j} \neq 0 \vee \ell^i_{\mathcal{I}_j-1} \in \mathcal{S})$.

	For each index $j \in \lbrace 1,2,\ldots,k-1\rbrace$, we output the commands \textit{inject}, \textit{run}, and \textit{store}
	by making use of the sequence $\mathcal{I}$ in this way.
	\begin{enumerate}
		\item If there is a disturbance to inject $d^i_{\mathcal{I}_{j}} \neq 0$, then $\text{{\tt print I}}d^i_{\mathcal{I}_{j}}$;
		\item $\text{{\tt print R}}(\mathcal{I}_{j+1}-\mathcal{I}_j)$;
		\item If the reached label $\ell^i_{\mathcal{I}_{j+1}-1}$ has to be stored,
			\textit{i.e.}, $\ell^i_{\mathcal{I}_{j+1}-1} \in \mathcal{S}$, then $\text{{\tt print S}}\ell^i_{\mathcal{I}_{j+1}-1}$,
			and remove it from $\mathcal{S}$, \textit{i.e.},
			$\mathcal{S} \gets \mathcal{S} \setminus \lbrace\ell^i_{\mathcal{I}_{j+1}-1}\rbrace$.
	\end{enumerate}
\end{description}

Note that the label $\ell^i_{\mathcal{I}_{j+1}-1}$ corresponds to the simulation interval reached by 
the previous command $run$. Also note that $p(i)$ is easily obtained from $\mathcal{L}_i$, and that $\ell^i_{j}$ is strictly identified 
by $i$ and $j$. Namely, $p(i) := \mathcal{L}_i\text{ }\%\text{ }H$, and $\ell^i_{j} := (i - 1) \cdot H + j$, for each $j=p(i)+1, \ldots, H$.

\clearpage
In conclusion, there is no need to load $\mathcal{D}$, $\mathcal{L}$, and $\mathcal{S}$ entirely into the main memory.
In fact, to generate the $i$-th line of simulator commands, all we need is the following data.
\begin{enumerate}[label=(\roman*)]
	\item The $i$-th DT $(d^i_{p(i)+1}, d^i_{p(i)+2}, \ldots, d^i_{H})$,
	\item The $i$-th label to load $\mathcal{L}_i$, and
	\item the first $H-p(i)$ elements in $\mathcal{S}$ to spot possible labels to store.\\
\end{enumerate}

\textbf{Example A.4.4} (Step 4). The example below shows the lex-order dataset of DTs $\mathcal{D}$,
the sequence of \textit{load labels} $\mathcal{L}$,
the lex-order sequence of distinct \textit{store labels} $\mathcal{S}$, and the final simulation campaign $\mathcal{C}$.

\begin{tcolorbox}[width=1.125\textwidth,colback=white,arc=0pt,bottom=5pt,boxrule=0.3pt]
\begin{Verbatim}[commandchars=\\\{\}]
\colorbox{gray!20}{Input 1} (Sorted DTs) \colorbox{gray!20}{Input 2} (Load Labels) \colorbox{gray!20}{Input 3} (Store Labels)

\colorbox{white}{0} \colorbox{white}{0} \colorbox{white}{0} \colorbox{white}{0} \colorbox{white}{0}        \colorbox{orange!60}{0}                     \colorbox{brown}{\textcolor{white}{1}}
\colorbox{white}{0} \colorbox{white}{0} \colorbox{blue!20}{1} \colorbox{white}{0} \colorbox{white}{0}        \colorbox{orange!60}{2}                     \colorbox{brown}{\textcolor{white}{2}}
\colorbox{white}{0} \colorbox{white}{0} \colorbox{blue!20}{1} \colorbox{blue!20}{2} \colorbox{white}{0}        \colorbox{orange!60}{8}                     \colorbox{brown}{\textcolor{white}{8}}
\colorbox{white}{0} \colorbox{blue!20}{1} \colorbox{white}{0} \colorbox{white}{0} \colorbox{blue!20}{1}        \colorbox{orange!60}{1}                     \colorbox{brown}{\textcolor{white}{23}}
\colorbox{blue!20}{1} \colorbox{white}{0} \colorbox{white}{0} \colorbox{white}{0} \colorbox{white}{0}        \colorbox{orange!60}{0}
\colorbox{blue!20}{1} \colorbox{white}{0} \colorbox{white}{0} \colorbox{blue!20}{1} \colorbox{white}{0}        \colorbox{orange!60}{23}

\colorbox{gray!20}{Output} (Optimised Simulation Campaign)

L\colorbox{orange!60}{0} R1 S\colorbox{brown}{\textcolor{white}{1}} R1 S\colorbox{brown}{\textcolor{white}{2}} R3
L\colorbox{orange!60}{2} I\colorbox{blue!20}{1} R1 S\colorbox{brown}{\textcolor{white}{8}} R2
L\colorbox{orange!60}{8} I\colorbox{blue!20}{2} R2
F2 F8 L\colorbox{orange!60}{1} I\colorbox{blue!20}{1} R3 I\colorbox{blue!20}{1} R1
F1 L\colorbox{orange!60}{0} I\colorbox{blue!20}{1} R3 S\colorbox{brown}{\textcolor{white}{23}} R2
L\colorbox{orange!60}{23} I\colorbox{blue!20}{1} R2
F23
\end{Verbatim}
\end{tcolorbox}

\clearpage
\section{Command-Line Tools}

In the following we describe the command line tools we implemented 
for the presented optimisation method. In particular, we first illustrate
each specific tool, and then we show how to use them 
to optimise an existing dataset of DTs.

\subsection{\tt dt-sort}

\begin{tcolorbox}[width=1\textwidth,colback=gray!10,arc=0pt,bottom=5pt,boxrule=0.2pt]
\begin{Verbatim}[commandchars=\\\{\}]
dt-sort [input DTs] [H] [B] [output DTs] (--unique)
\end{Verbatim}
\end{tcolorbox}

Sorts the input file of DTs that have horizon $H$.
Uses two buffers of size $B$ (bytes) to perform an IO-efficient external sorting based algorithm.
Removes duplicate DTs if the \texttt{--unique} argument is present.

\subsection{\tt dt-label}

\begin{tcolorbox}[width=1\textwidth,colback=gray!10,arc=0pt,bottom=5pt,boxrule=0.2pt]
\begin{Verbatim}[commandchars=\\\{\}]
dt-label [input sorted DTs] [H] [B] [output LL]
\end{Verbatim}
\end{tcolorbox}

Computes a file LL with \textit{load labels} from the input of uniquely-sorted DTs that have horizon $H$.
Uses one buffer of size $B$ to perform IO-efficient buffering of DTs.

\subsection{\tt dt-optimise}

\begin{tcolorbox}[width=1\textwidth,colback=gray!10,arc=0pt,bottom=5pt,boxrule=0.2pt]
\begin{Verbatim}[commandchars=\\\{\}]
dt-optimise [sorted DTs] [LL] [SL] [H] [B]
\end{Verbatim}
\end{tcolorbox}

Computes an optimised simulation campaign from:
(i)~a uniquely-sorted file of DTs that have horizon $H$,
(ii)~a file LL with \textit{load labels} that is obtained with \texttt{dt-label} from the same input DTs,
(iii)~a file SL with uniquely-sorted \textit{store labels} that is obtained with \texttt{dt-sort} from the LL input file.

Uses two buffers of size $B$ to perform IO-efficient buffering of the input files.
Prints the resulting simulation campaign on the standard output.

\clearpage
\subsection{\tt dt-merge}

\begin{tcolorbox}[width=1\textwidth,colback=gray!10,arc=0pt,bottom=5pt,boxrule=0.2pt]
\begin{Verbatim}[commandchars=\\\{\}]
dt-merge [DTs 1] [DTs 2] [H] [B] [output DTs] (--unique)
\end{Verbatim}
\end{tcolorbox}

Merges two input files with lex-ordered DTs.
These input DT files both have horizon $H$ and can contain
either sorted or uniquely-sorted DTs.

Uses two buffers of size $B$ to perform IO-efficient buffering of input DTs.
Removes duplicate DTs if the \texttt{--unique} argument is present.

\subsection[Usage]{Optimisation Example}

This is an example of how to use the command line tools described above.
In particular, we show how to perform each one of the four steps of our 
optimisation method.

Let {\tt in.DT} be an input file of DTs with horizon $H = 10$, and
let's use a number $B = 1000000$ of bytes per buffer.

\textbf{Step 1} (Initial Sorting). First, we use {\tt dt-sort} to compute the lex-ordered dataset $\mathcal{D}$ of DTs
in a file named {\tt in.DT.usort}.
\begin{tcolorbox}[width=1.05\textwidth,colback=white,arc=0pt,left=2pt,bottom=3pt,boxrule=0.2pt]
\begin{verbatim}
dt-sort in.DT 10 1000000 in.DT.usort --unique
\end{verbatim}
\end{tcolorbox}

\textbf{Step 2} (Load Labelling). Second, we use {\tt dt-label} to compute the file named {\tt in.DT.usort.LL}
with the sequence of {\em load labels}, starting from the lex-order dataset $\mathcal{D}$
computed at Step 1.
\begin{tcolorbox}[width=1.05\textwidth,colback=white,arc=0pt,left=2pt,bottom=3pt,boxrule=0.2pt]
\begin{verbatim}
dt-label in.DT.usort 10 1000000 in.DT.usort.LL
\end{verbatim}
\end{tcolorbox}

\textbf{Step 3} (Store Labelling). Third, we use {\tt dt-sort} again to compute the file named {\tt in.DT.usort.SL}
with the lex-order sequence of distinct {\em store labels}, starting from the sequence of {\em load labels}
computed at Step 2.
\begin{tcolorbox}[width=1.05\textwidth,colback=white,arc=0pt,left=2pt,bottom=3pt,boxrule=0.3pt]
\begin{verbatim}
dt-sort in.DT.usort.LL 1 1000000 in.DT.usort.SL --unique
\end{verbatim}
\end{tcolorbox}

In the command line above we have $H = 1$ (\textit{i.e.}, the 2nd argument of {\tt dt-sort}).
This is due to the fact that the input file with \textit{load labels}
is seen as a sequence of DTs with horizon~1.

\clearpage
\textbf{Step 4} (Final Optimisation). Last, we compute the final optimised simulation campaign starting
from the files computed at previous steps.
\begin{tcolorbox}[width=1.05\textwidth,colback=white,arc=0pt,left=2pt,bottom=3pt,boxrule=0.2pt]
\begin{verbatim}
dt-optimise in.DT.usort in.DT.usort.LL in.DT.usort.SL 10 1000000
\end{verbatim}
\end{tcolorbox}

Please note that since {\tt dt-optimise} prints the final simulation campaign on the standard output,
it may be desirable to redirect it into a file when dealing with large input datasets.